\documentclass[preprint,12pt,authoryear]{elsarticle}

\usepackage[utf8]{inputenc}
\usepackage[T1]{fontenc}
\usepackage{amsmath,amssymb}
\usepackage{graphicx}
\usepackage{booktabs}
\usepackage{longtable}
\usepackage{makecell}
\usepackage{ragged2e}
\usepackage{setspace}
\usepackage{textcomp} % for \pounds
\usepackage{siunitx}  % for \num{...}
\usepackage{placeins} % provides \FloatBarrier
\usepackage{hyperref}

\setcitestyle{authoryear,round}

\begin{document}
\justifying

\begin{frontmatter}

\title{The payment heterogeneity index: an integrated unsupervised framework for high-volume procurement oversight and decision support}

\author[aff1,aff2]{Kyriakos Christodoulides\corref{cor1}}
\cortext[cor1]{Corresponding author.}
\ead{kyriakos@novelintelligence.ai}

\affiliation[aff1]{organization={Philips University, Department of Computer Science}, city={Nicosia}, country={Cyprus}}
\affiliation[aff2]{organization={Novel Intelligence}, city={London}, country={UK}}

\begin{abstract}Public procurement is vulnerable to error, fraud, and corruption, particularly as high transaction volumes overwhelm oversight. While research often focuses on tender-stage anomalies, post-award payment monitoring remains underexplored. Since labelled datasets are rare and methods like Benford's Law face restrictive assumptions, there is a need for interpretable, unsupervised frameworks for high-volume procurement oversight and decision support. This paper introduces the Structural Heterogeneity Index (SHI), a composite statistic for one-dimensional samples, and its payment-specific instantiation, the Payment Heterogeneity Index (PHI), characterising payment structure and latent regimes. It incorporates Gaussian Mixture Model (GMM) parameters alongside non-parametric statistics, integrating four interpretable components: modality, asymmetry, tail behaviour, and structural dispersion. Uniquely, the tail-behaviour component captures both distributional heaviness and extreme-value concentration, while structural-dispersion combines the variability, prevalence, and separation of latent payment regimes. Applied to UK municipal procurement data, PHI identifies a financially significant cohort (0.6\% of suppliers; 10.1\% of high-volume vendors) with structurally distinct payment patterns. Statistical testing further supports these differences, and targeted human verification confirms the plausibility of prioritised cases. Comparative analysis shows PHI reveals regime separation obscured by the Coefficient of Variation ($\rho = 0.310$). PHI provides a transparent, decomposable, and computationally lightweight framework for procurement integrity oversight and targeted audit prioritisation.
\end{abstract}


\end{frontmatter}

\section{Introduction}

Public procurement represents a substantial portion of government spending. According to the Organisation for Economic Co-operation and Development (OECD), public procurement spending averages around 12--13\% of GDP and approximately 29--30\% of total government expenditure among member countries \citep{OECD2019}, making procurement integrity critical for fiscal accountability and economic efficiency. Public procurement is also widely recognised as vulnerable to waste, error, and corruption, motivating the development of practical monitoring tools and governance mechanisms \citep{OECD2016}. Within procurement and contract delivery processes, patterns in payment amounts, including splitting and clustering around focal values, can warrant scrutiny because they may be consistent with attempts to avoid controls or exploit discretion \citep{Spagnolo2012, Palguta2017, Coviello2018}. Recent work also highlights how discretion can create scope for favouritism and manipulation in procurement settings \citep{Szucs2024}. Related evidence also shows that the balance between formal rules and administrative discretion can materially affect procurement behaviour and outcomes \citep{Carril2021}. At the same time, similar patterns can arise from benign operational structure, such as fixed-fee arrangements, recurring billing schedules, heterogeneous service mixes, and administrative constraints or process frictions. Observed payment amounts reflect not only supplier-side billing regimes but also payment processes and controls on the part of the public body such as approval practices, batching, and settlement routines. Accordingly, complex distributional structure is treated in this study as a prioritisation signal (i.e., a procedure that ranks cases) for subsequent human review.

Existing literature on procurement risk detection has predominantly focused on tender-stage anomalies, such as bid rigging, single-bidder tenders, and collusive networks, using graph-based methods, econometrics, and supervised or semi-supervised machine learning approaches \citep{Caglayan2022, Rabuzin2019}. These models typically analyse features at the tender or contract level, including the number of bidders, award criteria, and supplier networks \citep{Nai2022, Carvalho2024}. While this work has advanced risk measurement at the award stage, important dynamics can occur after contract award during implementation. Post-award payment records capture actual disbursements of public funds, which can reveal structured payment regimes, shifting payment levels, or irregularities that merit follow-up. Despite this direct relevance to contract management and audit triage, where material impact is highest due to realised financial exposure, such post-award payment datasets, to our knowledge, remain understudied compared to tender-stage data in procurement analytics literature. 

In practice, auditors consider a range of indicators, including payment-amount regularities such as repeated identical values, and may also apply digit-based tests such as Benford's Law \citep{Nigrini2012}. Although Benford’s Law is an established diagnostic tool, its scope is limited to the distribution of leading digits, which may not capture the full range of atypical patterns relevant for targeted review. This motivates complementary unsupervised methods that move beyond digit patterns to capture broader distributional structure in a single interpretable measure. To our knowledge, no prior work has systematically operationalised such a multidimensional synthesis of morphological features for systemic cross-supplier benchmarking within the specific context of post-award public procurement data. Additionally, because confirmed labels for fraud, misconduct, or negligence are typically unavailable in procurement payments data, we adopt an unsupervised approach and prioritise interpretability by defining a score in terms of transparent, decomposable components with clear qualitative meaning, rather than optimising for predictive accuracy \citep{Ngai2011,West2016}. Recent work in data science and management has similarly emphasised the need to align analytical methods with sector-specific constraints and governance requirements in fraud and anomaly detection \citep{Chen2025}.

Gaussian Mixture Models (GMMs) provide a flexible and interpretable framework for modelling one-dimensional distributions as weighted combinations of Gaussian components, enabling the explicit characterisation of latent multimodality, spread, and component weights \citep{McLachlan2000,Chandola2009,Hodge2004}. While GMM-based approaches have been applied to tender-stage risk assessment \citep{Carvalho2024}, this study utilises GMM parameters as a foundational input within a broader synthesis of empirical distributional shape characteristics. The resulting decomposable, supplier-level heterogeneity score aids procurement management by enriching post-award monitoring and enabling systematic auditing prioritisation and resource allocation. 

Specifically, we introduce the Structural Heterogeneity Index (SHI) and its specialised application, the Payment Heterogeneity Index (PHI). The PHI operates on preprocessed transaction data to identify suppliers whose payment distributions deviate simultaneously across multiple structural dimensions: modality, asymmetry, tail behaviour and structural dispersion. By  aggregating these factors, the PHI captures the synergistic effect of distributional anomalies, rather than isolated irregularities, thereby focusing triage effort in workflows where prioritisation is critical. Beyond immediate audit utility, the PHI framework establishes a basis for long-term operational resilience. Specifically, it could translate complex distributional shifts into actionable KPIs for strategic supplier management and risk monitoring.

We demonstrate PHI using the City of York Council payments-to-suppliers dataset (2025–2026), a UK transparency release providing transaction-level payments above a £250 disclosure threshold \citep{CityYork2025}. This source is representative of administrative data generated under public transparency initiatives that mandate openness and accountability \citep{Janssen2012,Bertot2010}.

This paper evaluates PHI as an unsupervised discovery and prioritisation signal for high-volume post-award payment oversight. Three research questions guide the analysis. First, how can structural heterogeneity in supplier payment distributions be formalised as an interpretable composite of modality, robust asymmetry, tail behaviour, and structural dispersion? Second, does the structural dispersion component distinguish materially separated payment regimes from benign multi-tier payment structures that are merely multimodal? Third, when applied to real procurement payments, does PHI produce a small, interpretable, financially material shortlist that can support expert-led audit triage and motivate further field validation? By addressing these questions, the study presents PHI as a promising decision-support artefact for further mathematical, empirical, and operational evaluation.

The contribution is therefore threefold. First, the paper shifts attention from tender-stage indicators to post-award payment-regime structure, a comparatively underdeveloped area in procurement analytics. Second, it introduces SHI as a new composite statistic for summarising distributional heterogeneity in one-dimensional samples, defined as a function of four interpretable components: modality ($M$), asymmetry ($A$), tail behaviour ($T$), and structural dispersion ($D$). PHI is then defined as a specific multiplicative, decomposable instance of SHI suitable for per-supplier attribution. The novelty of SHI lies in its incorporation of a tail-behaviour component that remains sensitive to both extreme outliers and point clustering, and a structural-dispersion component ($D$) that distinguishes materially separated payment regimes from benign multimodal tiering. Third, it provides a proof-of-concept empirical assessment showing that PHI can generate a small, financially material, expert-reviewable shortlist for procurement oversight.

The remainder of this paper is organised as follows. Section~\ref{sec:data} details the data preparation and the GMM-augmented methodology used to identify latent operational regimes within payment flows. Section~\ref{sec:theory} develops the theoretical framework, formalising the Structural Heterogeneity Index (SHI) and justifying its specialised instantiation, the Payment Heterogeneity Index (PHI), as a synthesis of modal complexity and distributional shape. Section~\ref{sec:results} provides a proof-of-concept empirical assessment using real-world public procurement data; we deconstruct the PHI signal through expert-reviewed supplier cases and aggregate risk-tiering to assess its plausibility as a prioritisation signal for human-in-the-loop oversight and decision support. Furthermore, this section presents an exploratory analysis of a systemic characteristic termed \textit{threshold anchoring}, examining whether high-PHI payment modes show stronger interaction with discrete, systemically recurring financial values that are also visible in the wider cleaned dataset. Finally, Section~\ref{sec:discussion} discusses the strategic implications of these structural signals for procurement integrity, addresses methodological limitations, and outlines directions for future mathematical and empirical research.

\section{Data and Methodology}
\label{sec:data}

\subsection{Data Description and Context}

We analyse a six-month snapshot of the City of York Council ``Payments to Suppliers over \pounds 250'' dataset, covering May 2025 through October 2025 \citep{CityYork2025}. At the time of analysis, this UK local authority transparency release contained 38,966 payments totalling approximately \pounds 169 million across 1,911 unique creditor names. Each record includes: Organisation Name, Directorate, Department, Service Plan, Creditor Name (supplier), Payment Date, Transaction No, Card Transaction, Net Amount, Irrecoverable VAT, Subjective Group, Subjective Subgroup, and Subjective Detail. The unit of analysis is the supplier-level payment-amount distribution, derived from the \texttt{Net Amount} field. This cross-sectional approach provides a controlled signal for validating the initial framework; investigating the longitudinal stability of these signatures over longer fiscal cycles remains a subject for future research.

Although the dataset is labelled ``payments over \pounds 250,'' initial analysis reveals that the City of York Council publishes individual invoice-level payments for suppliers whose aggregate spend exceeds the disclosure threshold. Consequently, the dataset includes many transactions below \pounds 250 (minimum observed: \pounds 0.06; 2,642 payments between \pounds 0--\pounds 250). We therefore treat the disclosure rule as operating at the aggregate supplier level rather than the transaction level, allowing for a more granular analysis of payment regimes.

The data reflect realised payments rather than contracted amounts and therefore capture post-award implementation dynamics. As such, they may reflect both supplier-side billing regimes and public-body payment processes, such as approval practices, batching, and settlement routines. Accordingly, the analysis is designed to produce a practical, human-in-the-loop prioritisation signal by identifying suppliers whose payment-amount distributions exhibit atypical structural patterns, rather than to infer wrongdoing or intent.

To systematically operationalise supplier-level structure, we apply a pipeline consisting of: data cleaning and harmonisation, volume-based supplier selection, robust standardisation of amounts, GMM fitting with model selection, and PHI computation for prioritisation.

\subsection{Data Preparation and Methodological Pipeline}

\paragraph{Filtering Non-Standard Amounts}

We exclude non-positive net amounts (refunds, credits, and adjustments), removing 681 rows (1.75\% of observations) and leaving a final sample of 38,285 records. This step is necessary to isolate the joint structure of suppliers’ billing behaviour and the authority’s payment structuring, as well as to avoid statistical distortions introduced by accounting reversals. In a GMM setting, reversals can induce spurious, low-weight components and inflate variance, due to their low frequency and numerical distance from the main mass of transactions. While the presence of frequent credit notes is a known risk indicator, these records are typically linked to prior transactions; including them would partially conflate original billing events with subsequent adjustments and obscure the primary billing regimes we seek to model. We view the analysis of credit-note ratios as a valuable but separate diagnostic dimension to be explored in future work. Focusing exclusively on positive payment flows ensures that the PHI remains a targeted measure of the primary distributional profiles generated by standard billing and settlement cycles.

\paragraph{Identity Harmonisation and Ethics}

To prevent identity fragmentation, we apply a two-step fuzzy-harmonisation process. First, name strings are normalised by converting to uppercase, removing punctuation, and standardising common suffixes (e.g., ``LTD'' and ``LIMITED''). Second, we treat two strings as a match only if \emph{all} of the following conditions hold: Term Frequency--Inverse Document Frequency (TF--IDF) cosine similarity $\geq 0.76$, token set ratio $\geq 77$, token Jaccard overlap $\geq 0.36$, and an ensemble score $> 0.66$. These thresholds were calibrated via empirical inspection to be sufficiently conservative for this demonstration. 

Robust preprocessing is a necessary prerequisite for the PHI framework, as significant identity fragmentation or erroneous merging directly alters the observed shape complexity of a supplier's payment distribution. Following the matching stage, names are clustered using connected components in the match graph and assigned a canonical identifier based on the modal string in each cluster. This process reduces 1,911 raw names to 1,896 canonical suppliers. Table~\ref{tab:harmonisation_results} provides illustrative examples of the successfully merged variants. While exhaustive record linkage is a distinct research challenge, this empirical configuration was found to be sufficiently robust for the primary goal of this paper: demonstrating the diagnostic utility of the PHI framework. Future work may explore the integration of large language models for more granular name-cleansing workflows. Following this harmonisation stage, all supplier identities were pseudonymised to ensure compliance with GDPR and research ethics standards regarding the processing of personal data (e.g., sole traders).

\renewcommand{\arraystretch}{1.15}
\setlength{\tabcolsep}{6pt}

{\centering
\begin{longtable}{>{\RaggedRight}p{7.0cm} >{\centering\arraybackslash}p{2.0cm} >{\RaggedRight\arraybackslash}p{6.0cm}}
\caption{Summary of supplier name harmonisation results}
\label{tab:harmonisation_results} \\

\toprule
\textbf{Raw Names Merged} & \textbf{Number of Raw Names} & \textbf{Canonical Name} \\
\midrule
\endfirsthead

\multicolumn{3}{c}{{\bfseries \tablename\ \thetable{} -- continued from previous page}} \\
\toprule
\textbf{Raw Names Merged} & \textbf{Number of Raw Names} & \textbf{Canonical Name} \\
\midrule
\endhead

\midrule \multicolumn{3}{r}{{Continued on next page}} \\
\endfoot

\bottomrule
\endlastfoot

\makecell[l]{MUDDY BOOTS NURSERY\\ MUDDY BOOTS NURSERY LTD\\ MUDDY BOOTS NURSERY POPPLETON}
& 3
& \makecell[l]{MUDDY BOOTS NURSERY} \\
\midrule

\makecell[l]{ACOMB METHODIST CHURCH HALL\\ ACOMB METHODIST CHURCH}
& 2
& \makecell[l]{ACOMB METHODIST CHURCH \\HALL} \\
\midrule

\makecell[l]{BURTON GREEN PRIMARY SCHOOL\\ BURTON GREEN PRIMARY ACADEMY\\ SCHOOL}
& 2
& \makecell[l]{BURTON GREEN PRIMARY \\SCHOOL} \\
\midrule

\makecell[l]{E ON NEXT ENERGY\\ E ON NEXT ENERGY LTD}
& 2
& \makecell[l]{E ON NEXT ENERGY} \\
\midrule

\makecell[l]{HEATHCOTES CARE\\ HEATHCOTES CARE LTD}
& 2
& \makecell[l]{HEATHCOTES CARE} \\
\midrule

\makecell[l]{MOTUS COMMERCIALS LTD\\ MOTUS COMMERCIALS}
& 2
& \makecell[l]{MOTUS COMMERCIALS LTD} \\
\midrule

\makecell[l]{NEW EARSWICK PRIMARY SCHOOL\\ NEW EARSWICK PRIMARY ACADEMY}
& 2
& \makecell[l]{NEW EARSWICK PRIMARY \\SCHOOL} \\
\midrule

\makecell[l]{OUR LADY QUEEN OF MARTYRS RC VA \\PRIMARY SCHOOL\\ OUR LADY QUEEN OF MARTYRS RC \\PRIMARY ACADEMY\\}
& 2
& \makecell[l]{OUR LADY QUEEN OF MARTYRS \\RC VA PRIMARY SCHOOL} \\
\midrule

\makecell[l]{POPPLETON ROAD PRIMARY SCHOOL\\ POPPLETON ROAD PRIMARY \\ACADEMY SCHOOL}
& 2
& \makecell[l]{POPPLETON ROAD PRIMARY \\SCHOOL} \\
\midrule

\makecell[l]{SUNSHINE DAY NURSERY\\ SUNSHINE DAY NURSERY YORK LTD}
& 2
& \makecell[l]{SUNSHINE DAY NURSERY} \\
\midrule

\makecell[l]{WEST THORPE PRE SCHOOL\\ WEST THORPE PRE SCHOOL \\PLAYGROUP}
& 2
& \makecell[l]{WEST THORPE PRE SCHOOL} \\
\midrule

\makecell[l]{YORK CITY FOOTBALL CLUB \\FOUNDATION\\ YORK CITY FOOTBALL CLUB}
& 2
& \makecell[l]{YORK CITY FOOTBALL CLUB \\FOUNDATION} \\
\end{longtable}
\par}

\paragraph{Selecting Suppliers with Sufficient Payments}

We restrict the analysis to suppliers with at least 50 payments during the observed period, yielding a subset of 119 suppliers and 26,721 records (69.8\% of positive-amount observations). This threshold ensures sufficient data density for reliable distributional characterisation and positions the PHI as a specialised tool for prioritising high-frequency, recurring suppliers. While the six-month window represents a partial fiscal cycle, the $n \ge 50$ requirement provides a sufficient basis for estimating stable distributional signatures. Low-frequency suppliers are excluded from this specific framework not because they represent lower risk, but because they are better addressed through complementary controls, such as rule-based checks or manual review, where mixture-based structural estimates would be statistically unreliable.

\paragraph{Robust Standardisation}

To enable cross-supplier comparison while reducing sensitivity to extreme values, payment amounts $a_i$ are standardised using the \emph{global} median $med(A)$ and \emph{global} interquartile range (IQR), computed over all payments in the dataset:
\[
\tilde{a}_i = \frac{a_i - med(A)}{\mathrm{IQR}}, \qquad \mathrm{IQR}=Q_3-Q_1.
\]
Global, rather than within-supplier, scaling places all suppliers on a common monetary reference frame while preserving relative distances. This ensures that between-mode separations reflect financially meaningful differences for prioritisation, as larger separations in the standardised space correspond to larger absolute amounts at stake.

Note that because robust standardisation is performed using a global median and interquartile range, the transformed amounts naturally take both positive and negative values. Negative values simply indicate payments below the global median and carry no intrinsic meaning beyond relative position. All downstream PHI components are scale-free and depend on relative distances, quantiles, or absolute separations; therefore, the presence of negative values does not affect interpretability or comparability across suppliers.

\begin{figure}[htbp]
    \centering
    \includegraphics[width=\textwidth,height=\textheight,keepaspectratio]{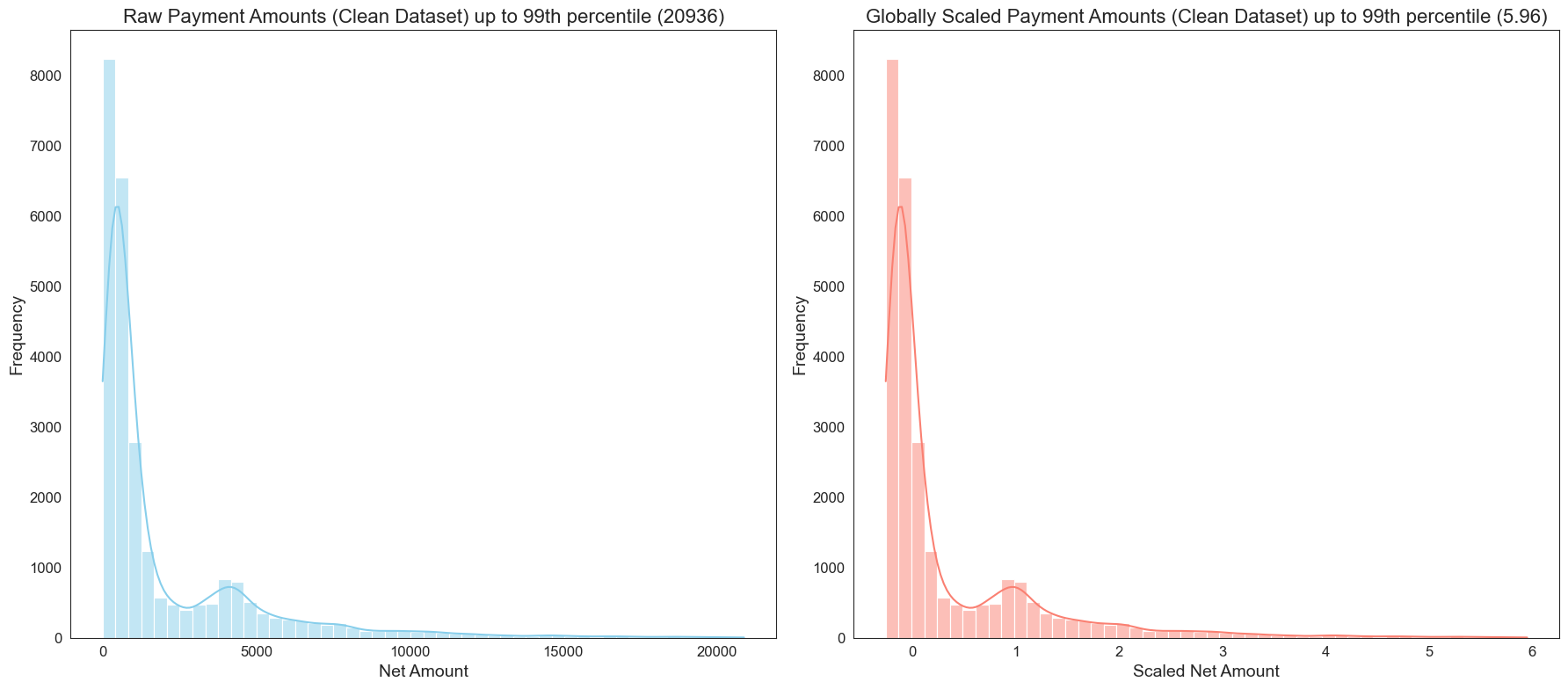}
    \caption{Distribution of robustly standardised payment amounts, shown as a histogram with kernel density overlay (truncated at the 99th percentile for visual clarity).}
    \label{fig:robust_standardisation}
\end{figure}

Figure~\ref{fig:robust_standardisation} illustrates the effect of this transformation, showing histograms with kernel density estimate overlays truncated at the 99th percentile for visual clarity.

\paragraph{Gaussian Mixture Models for PHI Operationalisation}

We employ Gaussian Mixture Models (GMMs) to characterise the underlying payment regimes (modes) of each distribution; for visual exposition, these structural clusters are occasionally referred to as activity zones. Separately, non-parametric statistics of each supplier's full payment distribution are used as  robust features in the PHI calculation. A GMM represents a distribution as a weighted sum of Gaussian components, each defined by parameters $(\mu_i, s_i, \pi_i)$ estimated via the Expectation-Maximization (EM) algorithm\citep{Dempster1977}.  To control model complexity and minimise the users' cognitive load, we cap the number of mixture components. A univariate $k$-component GMM has $3k-1$ free parameters (means, variances, and $k-1$ independent weights). Following a conservative heuristic of approximately 20--25 observations per free parameter \citep{McLachlan2000}, we set the maximum number of components as:
\[
k_{\max}=\min\left(4, \left\lfloor \frac{n}{25} \right\rfloor\right).
\] 
For each supplier, we fit GMMs with $k \in \{1, \dots, k_{\max}\}$ and select the model with the lowest Bayesian Information Criterion (BIC) score \citep{Schwarz1978}. This constraint ensures that a supplier requires at least 50 transactions to fit a 2-component model (5 parameters) and approximately 100 transactions to consider a 4-component model (11 parameters). Additionally, capping the maximum at $k=4$ aims to identify systematic payment regimes rather than fragmenting the distribution into overly granular, idiosyncratic, or extreme-value clusters.

Complementing this cap and to reinforce that the index captures meaningful structural regimes, we exclude components with negligible weight ($\pi_i < 0.05$) and renormalise the remaining weights to sum to unity. While higher-order mixtures or alternative parameterisations may be statistically viable, we adopt this constrained framework under conservative assumptions, to establish a stable and interpretable foundation for the proposed index. As this work represents an initial proof-of-concept evaluation of PHI, we treat these settings as transparent operational defaults rather than optimised parameters. Systematic hyper-parameter sensitivity and cross-institutional calibration are important extensions for future work.

We implement supplier‑level GMMs using scikit‑learn's \texttt{GaussianMixture} \citep{Pedregosa2011}, setting \texttt{covariance\_type='spherical'}, which is consistent with the single-variance parameterisation of our univariate model. For reproducibility we fix the EM hyperparameters: the initialisation scheme \texttt{init\_params} is set to k‑means, the number of random restarts \texttt{n\_init} is set to $1$, the convergence tolerance \texttt{tol} to $10^{-3}$, the maximum number of EM iterations \texttt{max\_iter} to $M=100$, and the RNG seed by $s_{\text{seed}} = 0$. A small covariance regulariser $\lambda=10^{-6}$ is applied uniformly to prevent pathological component collapse; this stability safeguard is constant across suppliers, not tuned to the data, and does not materially affect regime identification or component separation.

We emphasise that we employ GMMs as a parametric device to identify and summarise distinct payment regimes, focusing on modal structure and component separation, rather than as a literal model of the underlying distribution. This design choice is motivated by the practical reality of procurement data: sparse distributions yield less stable estimates than high-volume ones, a limitation acknowledged in the design of the frequency threshold.       

\paragraph{Risk Ranking and Application of PHI Scores}

To facilitate practical use, raw PHI scores are converted into percentile ranks within the analysed supplier population. This percentile-based ranking provides a familiar 0--100 scale, enabling straightforward tiering and resource prioritisation without obscuring the underlying modelling approach.

\paragraph{Operational Mapping of PHI Scores}

As a final step in the methodology, percentile-ranked PHI scores are mapped to qualitative risk tiers to support prioritisation. In a generic configuration, lower-percentile suppliers correspond to baseline monitoring, mid-range suppliers to enhanced review, and high-percentile suppliers to targeted audit attention. This mapping is intended as a flexible interpretive layer rather than a fixed decision rule; specific thresholds and associated actions depend on organisational context and risk tolerance.

\section{Theoretical Framework}
\label{sec:theory}

\subsection{Structural Heterogeneity Index (SHI)}

This paper proposes the Structural Heterogeneity Index (SHI) as a flexible, component-based framework for summarising the distributional structure in one-dimensional samples. While anomaly detection typically categorises deviations as point, contextual, or collective anomalies \citep{Chandola2009}, these methods often focus on identifying specific observations that deviate from a central tendency. In contrast, SHI is designed to capture the structural complexity of the underlying distribution itself. Rather than identifying individual outliers, SHI summarises a sample’s structural profile, enabling consistent and interpretable comparison across entities.

SHI is defined generally as an aggregation of four interpretable dimensions:

\begin{equation}
\text{SHI} = f(M, A, T, D)
\label{eq:shi_conceptual}
\end{equation}

where the mapping $f(\cdot)$ specifies whether these dimensions interact as independent additive risks or as mutually amplifying factors. In this paper, by introducing the PHI, we instantiate $f(\cdot)$ as a multiplicative function to reflect the principle of joint amplification, whereby the co-occurrence of multiple structural anomalies is weighted more heavily, and therefore prioritised relative to isolated structural deviations.  Additionally, the multiplicative instantiation of PHI permits the exact additive decomposition of $\ln(\mathrm{PHI})$ into per-component log-contributions and therefore supports per-supplier attribution. 

The SHI framework is inherently modular and depending on oversight objectives or data availability, practitioners may choose to include or omit individual (sub-)components. The structural components are computed for each distinct sample or entity-level distribution and represent:

\begin{itemize}
    \item $M$ (\textbf{Modality}): Captures the number of distinct modes within the distribution. Higher values may indicate fragmented structural processes (e.g., ``Does the entity operate across multiple latent states or behaviours?'').

    \item $A$ (\textbf{Asymmetry}): Measures systematic directional bias or skew. It distinguishes between balanced variations and those concentrated in a specific direction (e.g., ``Is there an inherent directional bias in the observed events?'').

    \item $T$ (\textbf{Tail Behaviour}): Quantifies the concentration of mass in the extremes relative to the centre, remaining sensitive to both outliers and extreme point-clustering or ``heaping'' (e.g., ``Are extreme outcomes disproportionately represented or concentrated?'').

    \item $D$ (\textbf{Structural Dispersion}): Evaluates the combined structural variability, prominence and distinctness of the identified modes by considering both their internal scale (weight and dispersion) and their relative divergence from the dominant component. This is the key mode-architecture component of the index: it distinguishes distribution with multiple but closely spaced modes from distributions whose secondary components are materially separated from their dominant mode (e.g., ``Are there modes that are both prominent and distinct from the dominant mode?'').
\end{itemize}

The following section details the specific operationalisation of these structural dimensions for the analysed supplier payment dataset.

\subsection{Payment Heterogeneity Index (PHI): Robust Instantiation}

Building on the modelling framework described above, the PHI combines GMM-derived regime characteristics with robust quantile-based statistics to operationalise the structural dimensions. Specifically, for each supplier, given a standardised payment distribution $\tilde{\mathbf{A}}$, we fit a GMM with $k$ components defined by weights $\pi_i$, means $\mu_i$, and standard deviations $s_i$ and compute the corresponding non-parametric quantile statistics. All quantiles $Q_p$ in the following definitions are computed from the supplier-level distribution $\tilde{\mathbf{A}}$.

Let $i^\star = \arg\max_i \pi_i$ denote the dominant component index. The structural components are operationalised as follows:

\begin{enumerate}
    \item \textbf{Modality ($M$):} Defined as $M = k$. Here, $M=1$ represents the baseline unimodal state, while $M > 1$ acts as an amplifying factor for structural complexity.
    \item \textbf{Asymmetry ($A$):} Using Bowley skewness ($a_q$) for outlier resistance, we define $A = 1 + |a_q|$, where:
    \begin{equation}
     a_q = \begin{cases} 
     \frac{Q_{0.75} + Q_{0.25} - 2Q_{0.50}}{Q_{0.75} - Q_{0.25}}, & \text{if } Q_{0.75} \neq Q_{0.25} \\
     0 & \text{otherwise}
     \end{cases}
     \end{equation}
    \item \textbf{Tail Behaviour ($T$):} To capture both heavy tails and extreme concentration (heaping), we define $T = 1 + |\ln(t_q)|$, where:
    \begin{equation}
    t_q = \frac{(Q_{0.95}-Q_{0.05})+\varepsilon}{(Q_{0.75}-Q_{0.25})+\varepsilon}.
    \end{equation}
    The constant $\varepsilon$ ($10^{-6}$) ensures numerical stability under extreme price-point clustering.
    \item \textbf{Structural Dispersion ($D$):} Evaluates the structural dispersion of the identified regimes by combining (i) the weighted dispersion of the dominant regime and (ii) the weighted dispersion of secondary regimes, scaled by their distance ($d_i$) from the dominant regime. $D$ is not a fitting criterion; it is a post-hoc summary of the estimated mixture 
architecture.
    \begin{equation}
     D = 1 + \pi_{i^\star} s_{i^\star} + \sum_{i \neq i^\star} \pi_i s_i \ln(1 + d_i)
    \end{equation}
    where $d_i = |\mu_i - \mu_{i^\star}|$. The logarithmic term acts as a dampening function, ensuring that $D$ reflects the presence of materially distinct structural regimes, rather than being disproportionately driven by extreme distances associated with low-weight components (i.e., point-like outlier effects). The role of $D$ is therefore to assign structural significance to the regimes identified by the GMM. A supplier with several low-variance and tightly clustered regimes may have high modality but low structural dispersion, indicating tiered but operationally coherent payment behaviour. Conversely, a supplier with only two regimes can receive a high $D$ value if the secondary regime is both sufficiently weighted and materially separated from the dominant regime. In the unimodal case, $D$ can also increase when the dominant regime itself exhibits substantial dispersion, although this reflects broad within-regime variability rather than separation between distinct regimes. This property is important for audit triage because it down-weights benign multi-tier structures while elevating payment architectures consistent with different contractual scales, billing logics, or service regimes.
\end{enumerate}

The PHI is the multiplicative product: 
\begin{equation}
\text{PHI} = M \times A \times T \times D
\label{eq:phi_total}
\end{equation} 
Unlike additive models, coefficientless multiplication naturally prioritises concurrent anomalies across dimensions without requiring heuristic weight calibration. This ensures the index remains portable and fully decomposable for audit-trail explanations (e.g., identifying a high score as being driven specifically by multi-modality and tail volatility). Consistent with the operational pipeline, raw PHI values are converted into percentile ranks for organisational tiering and decision support.

\section{Results}
\label{sec:results}

\subsection{Sample Characteristics and Spend Concentration}

This section evaluates the Payment Heterogeneity Index (PHI) from four complementary angles. We begin by describing the analytical sample and the concentration of financial activity among high-frequency ( $n \ge 50$) suppliers. We then examine the distribution of PHI scores and introduce a risk-based tiering aligned with audit prioritisation. Next, we use illustrative case studies to exhibit how the PHI signal can be interpreted at the supplier level. Finally, we analyse the structural patterns associated with high PHI scores, also investigating whether payment regimes converge toward shared financial thresholds. Taken together, these results demonstrate both the operational utility and the prioritisation transparency of PHI as a decision-support tool.

\begin{figure}[htbp]
    \centering
    \makebox[\textwidth][c]{%
        \includegraphics[width=1.05\textwidth]{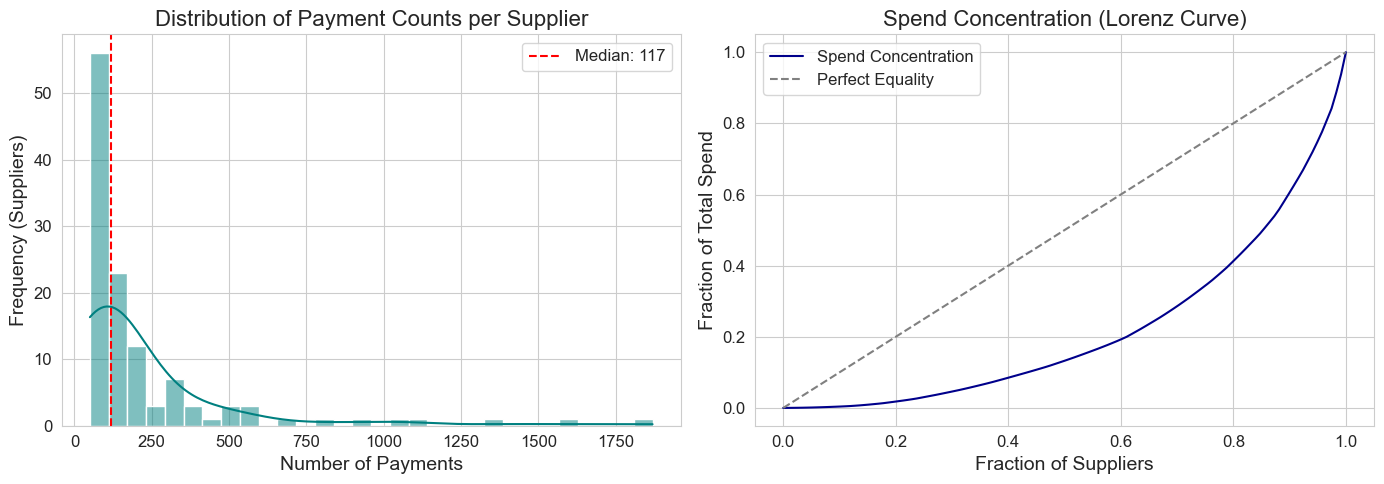}%
    }
    \caption{Analytical sample characteristics: Distribution of payment counts per high-frequency supplier (left) and Lorenz curve of spend concentration (right), highlighting the financial weight of a small subset of the filtered suppliers.}
    \label{fig:lorenz}
\end{figure}

The final analytical sample, following the application of the $n \ge 50$ volume threshold, includes \num{119} unique canonical suppliers. As shown in Figure~\ref{fig:lorenz} (left), these suppliers account for a median of \num{117} payments each (IQR: \num{67.5}--\num{223}), with a long tail extending to over \num{1750} transactions for the most active entities. This high transactional density provides the robust statistical basis required for Gaussian Mixture Model (GMM) estimation and the subsequent calculation of the Payment Heterogeneity Index (PHI). Individual payment amounts within this cleansed sample range from \pounds\,\num{0.06} to \pounds\,\num{702533.27}, with a median transaction value of \pounds\,\num{721.00}.

At the aggregate level, this subset represents the ``operational core'' of the authority. While it accounts for \num{39.40}\% of the Council's total \emph{positive} financial expenditure during the disclosure period, it captures \num{69.79}\% of all individual payment records. This concentration confirms that the sample is dominated by high-frequency, recurring payment activity, providing a suitable basis for identifying structural patterns. In contrast, low-frequency transactions, which drive the remaining \num{60.60}\% of spend, while potentially material, are less amenable to distributional analysis and are better addressed through complementary controls.

To visualise spend concentration within the sample, Figure~\ref{fig:lorenz} presents a Lorenz curve, plotting the cumulative share of total expenditure against the cumulative share of high-frequency suppliers. The pronounced deviation from the line of equality indicates that even among these recurring suppliers, financial impact is highly concentrated: the top \num{20}\% of suppliers in this group account for approximately \num{60}\% of the sample's spend. This suggests that the Council's operational dependencies are concentrated in a ``vital few'' entities where repeated anomalies could aggregate into significant financial loss or signal systemic process failure.

This level of concentration empirically reinforces the need for systematic prioritisation within the dataset, as structurally atypical behaviour among financially significant suppliers could carry substantial material impact. Although a small subset of high-frequency suppliers carries disproportionate financial weight, prioritisation is applied uniformly across the cohort based on structural characteristics rather than financial magnitude. The PHI thus segments suppliers into audit-prioritisation tiers solely according to payment structure, ensuring that prioritisation is driven by distributional atypicality.

\FloatBarrier

\subsection{Distribution of PHI Scores and Risk Tiering}

\begin{figure}[htbp]
    \centering
    \includegraphics[width=\textwidth,height=0.4\textheight,keepaspectratio]{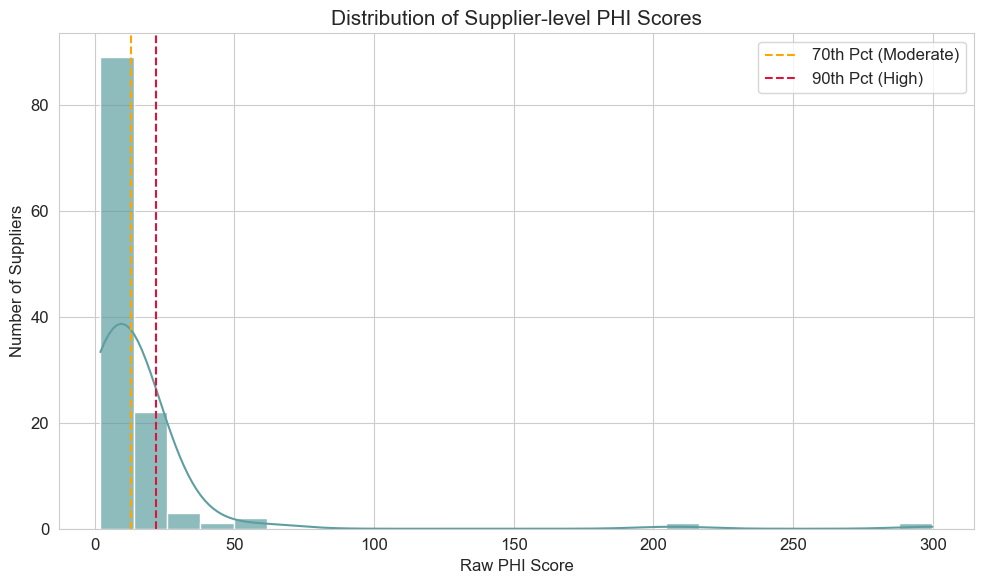}
    \caption{Distribution of supplier-level PHI scores. Vertical dashed lines indicate the 70th (orange) and 90th (red) percentile thresholds used for risk-based tiering. The pronounced right tail reflects scores extending to nearly an order of magnitude above the 90th percentile threshold.}
    \label{fig:PHI_Distribution}
\end{figure}

Applying the PHI framework to the analytical sample reveals a highly skewed distribution of structural complexity (Figure~\ref{fig:PHI_Distribution}). This empirical distribution supports the segmentation of suppliers into risk-based tiers; however, the specific percentile thresholds reflect an operational calibration aligned with the organisation’s risk appetite and audit capacity. As such, the tiering should be interpreted as a decision-support layer rather than a fixed classification, and may be adjusted to reflect alternative governance priorities or audit resource constraints.

\begin{itemize}
    \item \num{69.7}\% Low PHI (Standard oversight; 83 suppliers),
    \item \num{20.2}\% Moderate PHI (Thematic monitoring; 24 suppliers),
    \item \num{10.1}\% High PHI (Targeted audit triage; 12 suppliers).\footnote{The High‑PHI tier is defined as the top decile (\(\geq\)90th percentile). In an analytical sample of \num{119} suppliers the 90th percentile corresponds to 11.9 suppliers; to retain whole entities we select the top \num{12} suppliers, yielding $\tfrac{12}{119} \approx 10.1\%$. Tier boundaries for the other categories are determined analogously by mapping percentile cutoffs to whole supplier counts.}
\end{itemize}

The financial significance of the high-PHI tier is substantial. Although these entities represent only the top decile of suppliers by structural complexity, they account for \num{23.5}\% of total expenditure within the analytical sample. In the context of the full dataset, this corresponds to approximately \num{9.3}\% of the Council's total disclosed spend. This indicates that, in this dataset, structural atypicality is not confined to low-value suppliers; rather, it is present among entities handling significant financial volumes within the operational core of the authority.

We also note that the 90th percentile threshold, marking the start of the high-PHI tier, sits at approximately a score of \num{25}, while the distribution extends to extreme outliers with scores exceeding \num{200}. Within the High-PHI tier itself, there is nearly an order of magnitude difference between the lowest and highest scores, indicating substantial heterogeneity even among the highest-risk suppliers. This pronounced separation between the central mass and the extreme tail enables oversight teams to bypass routine variations and focus resources on the relatively small number of suppliers whose payment distributions exhibit the highest levels of structural complexity. Finally, PHI scores are inherently relative to the analysed population; percentile thresholds are not intended to be directly transferable across datasets without recalibration.

\subsection{Illustrative Supplier Case Studies}

To demonstrate the practicality and interpretability of the PHI as a prioritisation mechanism, we present five expert-reviewed supplier cases from the cleansed dataset. These cases were examined by the author using forensic accounting judgement informed by prior experience as a Certified Fraud Examiner. They should therefore be interpreted as expert-led plausibility checks of the PHI signal rather than confirmed legal or disciplinary findings. The `activity zones' of each case are modelled using a Gaussian Mixture Model (GMM), where the continuous profile (red curve) represents the fitted density and vertical dashed lines indicate the mathematical centres ($\mu_i$) of the identified payment regimes. With the exception of the maximum-PHI case, included as the primary audit target, the remaining cases were selected from each modality group (one to four modes) as indicative examples to evaluate whether the PHI behaves as intended across diverse distributional forms. These cases serve to showcase how a human-in-the-loop analyst might interpret specific component contributions and perform the subsequent forensic drill-downs required for atypical payment discovery. We reiterate that the PHI does not perform diagnosis; rather, it structures and prioritises suppliers for further investigation. Crucially, the following cases highlight that the PHI does not reward modality alone; rather, structural dispersion acts as a governing factor that distinguishes materially separated regime shifts from closely spaced, operationally intertwined variations.

To better understand what drives the prioritisation signal, we decompose the PHI score into its constitutive factors. This allows us to quantify the relative contribution of asymmetry, tail behaviour, modality, and structural dispersion. Because PHI is defined multiplicatively, the contribution of any single component $X \in \{M, A, T, D\}$ to the overall score is naturally expressed on the log scale:
\begin{equation}
\text{Contribution}_X = \left( \frac{\ln(X)}{\ln(\text{PHI})} \right) \times 100\%
\label{eq:contribution}
\end{equation}

All component percentages reported hereafter refer to this measure unless stated otherwise.

Table~\ref{tab:archetype_components} reports the resulting decomposition alongside the overall PHI scores for the selected suppliers. This provides a transparent link between the observed distributional structure and the resulting prioritisation level. The cases also illustrate the distinctive role of structural dispersion: Supplier C and Supplier D show that additional modes do not automatically imply high priority when regimes are closely spaced, whereas Supplier B and the maximum-PHI case show how fewer but materially separated regimes can generate stronger triage signals.

\begin{table}[htbp]
\centering
\caption{PHI components, contributions, scores, percentile ranks, and risk tiers for illustrative supplier cases}
\label{tab:archetype_components}
\resizebox{\textwidth}{!}{%
\begin{tabular}{lccccccc}
\toprule
Supplier & Modality (M) & Asymmetry (A) & Tail Behaviour (T) & Structural Dispersion (D) & PHI Score & Percentile Rank & Risk Tier \\
\midrule
Supplier A & 1.00 (0.0\%) & 1.08 (10.7\%) & 1.84 (83.1\%) & 1.05 (6.2\%) & 2.084 & 0.8 & Low \\
Supplier B & 2.00 (24.4\%) & 1.70 (18.7\%) & 2.14 (26.7\%) & 2.36 (30.2\%) & 17.150 & 83.2 & Moderate \\
Supplier C & 3.00 (57.1\%) & 1.32 (14.6\%) & 1.61 (24.6\%) & 1.07 (3.7\%) & 6.854 & 34.5 & Low \\
Supplier D & 4.00 (53.3\%) & 1.62 (18.4\%) & 2.08 (28.1\%) & 1.01 (0.2\%) & 13.494 & 72.3 & Moderate \\
Max PHI    & 2.00 (12.2\%) & 1.22 (3.4\%) & 4.43 (26.1\%) & 27.85 (58.3\%) & 299.836 & 100.0 & High \\
\bottomrule
\end{tabular}%
}
\end{table}

Moderate- and high-PHI suppliers together comprise a relatively small subset of the entire sample. However, the cases presented below illustrate that even moderate-PHI suppliers exhibit structural features that may merit ongoing monitoring, supporting the role of PHI as a graduated prioritisation tool rather than a binary risk classifier.

\paragraph{Supplier A: Unimodal Payment Structure}
Supplier A provides a representative example of a unimodal payment structure. The payment distribution is captured by a single-component GMM centred at $\mu = £487$ ($\pi = 1.00$), where payment amounts are smoothly distributed around this central tendency without structural fragmentation or multi-regime clustering. From a management perspective, this profile represents a high-frequency supplier characterised by predictable payment behaviour. While the PHI framework successfully detects a tail behaviour contribution ($83.1\%$ of its very low log-signal), this profile would be assigned a low prioritisation level in terms of structural complexity. Indeed, a granular inspection of the underlying transactions confirms this interpretation: all payments are categorised under `Services', with a narrow range of $£627$ and a median of $£468.75$, closely aligning with the GMM-identified Centre.
\begin{figure}[htbp]
    \centering
    \includegraphics[width=\textwidth,height=0.4\textheight,keepaspectratio]{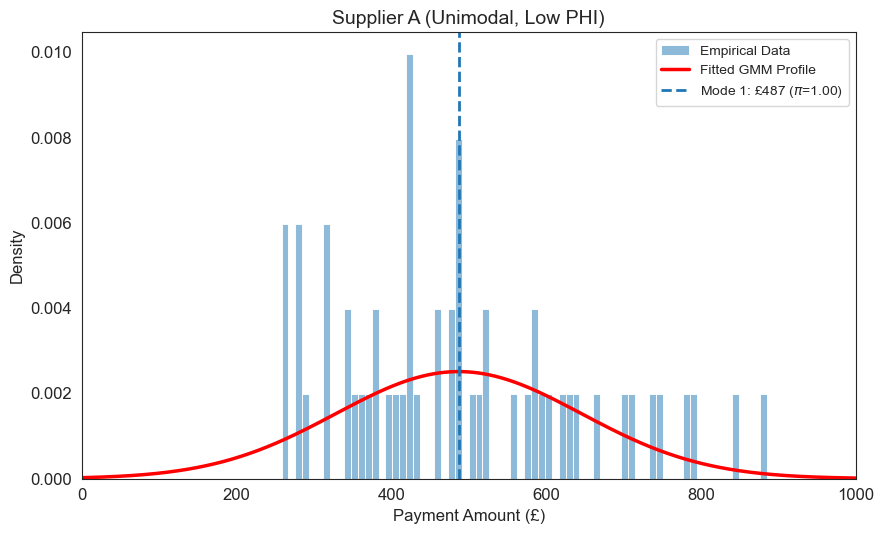}
    \caption{GMM profile for Supplier A. A single dominant payment regime is observed, with transactions concentrated around a central value and no evidence of structural fragmentation, consistent with a low prioritisation profile. Note: The x-axis scale differs across panels to preserve visibility of supplier-specific distributional structure.}
    \label{fig:supplierA}
\end{figure}

\paragraph{Supplier B: Bimodal Separation}
Supplier B exhibits a bifurcated payment structure characterised by two distinct activity zones: a high-frequency operational mode centred at $\mu_1 = £1{,}210$ ($\pi_1 = 0.52$) and a secondary, higher-value regime at $\mu_2 = £8{,}967$ ($\pi_2 = 0.48$). Although the PHI score is categorised as ``moderate,'' this supplier lies within the upper 16.8\% of PHI scores, indicating a clear structural departure from unimodal behaviour and positioning Supplier B as a borderline case between routine thematic monitoring and the targeted audit triage reserved for the highest-priority entities.

\begin{figure}[htbp]
    \centering
    \includegraphics[width=\textwidth,height=0.4\textheight,keepaspectratio]{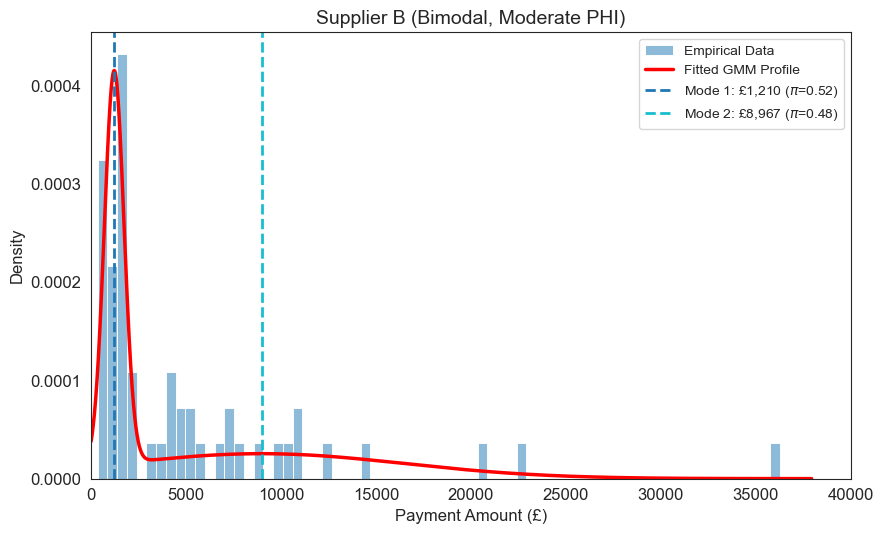}
    \caption{GMM profile for Supplier B. Two distinct payment regimes are identified, with a clear separation between a high-frequency operational cluster and a secondary higher-value regime. The bimodal structure and the presence of high-value outliers are consistent with the elevated Structural Dispersion and Tail Behaviour components that drive this 
supplier's PHI score. Note: The x-axis scale differs across panels to preserve visibility of supplier-specific distributional structure.}
    \label{fig:supplierB}
\end{figure}

\begin{figure}[p]
    \centering
    \includegraphics[width=\textwidth,height=0.4\textheight,keepaspectratio]{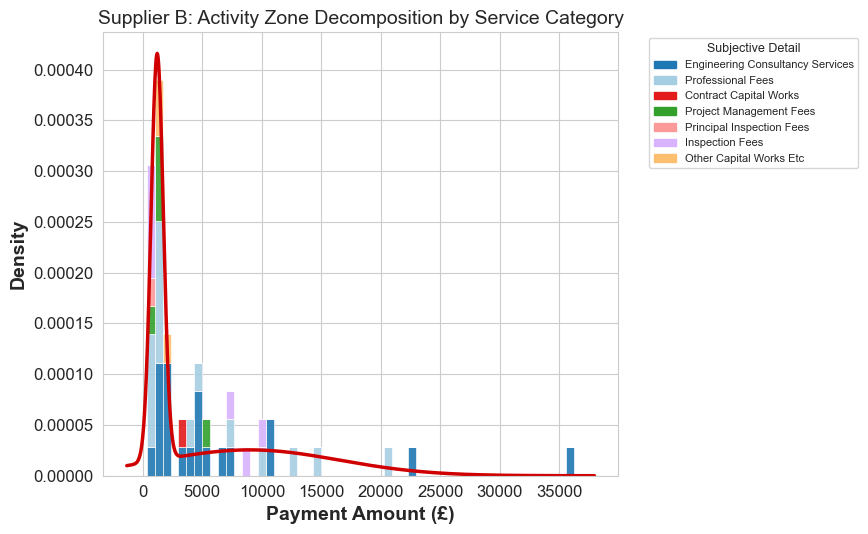}
    \caption{Probabilistic activity zone decomposition for Supplier B, showing the service categories driving the two latent payment regimes. The stacked bars indicate the relative contribution of each service category to the empirical density, with the fitted GMM profile overlaid in red.}
    \label{fig:supplierB_activity}
    \includegraphics[width=\textwidth,height=0.4\textheight,keepaspectratio]{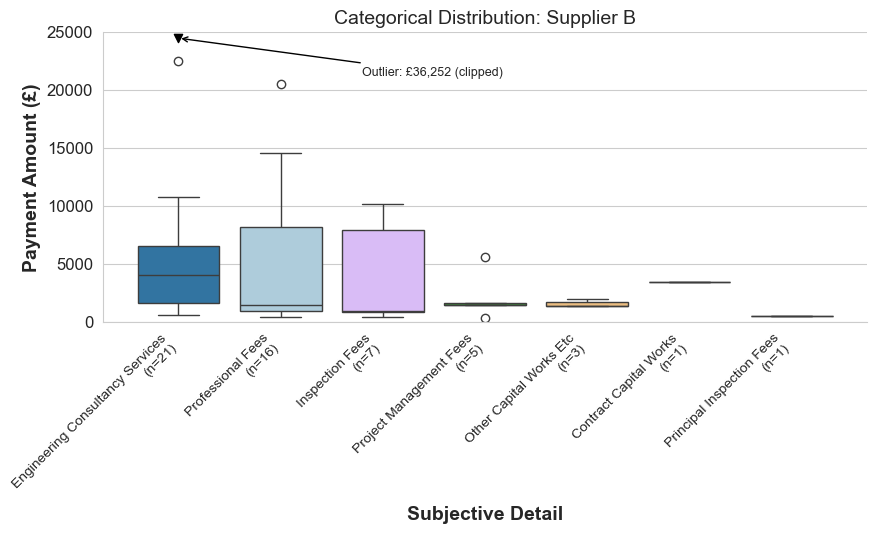}
    \caption{Categorical boxplot for Supplier B, showing the financial scale, median tendencies, and extreme outliers across service categories. The pronounced spread within Engineering Consultancy Services and Professional Fees provides categorical-level evidence supporting the elevated PHI signal.}
    \label{fig:supplierB_boxplot}
\end{figure}

The PHI signal for this supplier, is driven by three primary drivers (Equation~\eqref{eq:contribution}). Modality (24.4\%) identifies two distinct payment centres; Structural Dispersion (30.2\%) captures the substantial financial distance between the standard £1k service fees and the £9k cluster, as well as the variance within those groupings; and Tail Behaviour (26.7\%) reflects extreme outliers ranging as high as £36,251. To illustrate the human-in-the-loop interpretability of the PHI signal, Figures~\ref{fig:supplierB_activity} and~\ref{fig:supplierB_boxplot}  provide a categorical decomposition of Supplier B's payment distribution. These visualisations are not outputs of the PHI framework itself, but represent the type of targeted drill-down that a PHI signal is designed to motivate.

Inspection of the activity‑zone colouring and category boxplots reveals that the first, narrower peak comprises a dense mixture of Engineering Consultancy Services, Professional Fees, and Project Management Fees. Conversely, while the second, broader peak is also dominated by Engineering Consultancy and Professional Fees, it uniquely includes Contract Capital Works and a higher concentration of Inspection Fees. 

A key observation is the scaling of professional service fees into a higher-value regime. For example, while identical $£1{,}513$ entries populate the first mode, the second mode contains disbursements for the same Professional Fees category exceeding $£10{,}033$. This pattern represents the type of structural split that warrants targeted audit review to determine whether the higher-value regime is contractually justified. Specifically, an auditor would verify whether high-value payments categorised as Engineering Consultancy Services correspond to pre-authorised capital milestones, distinct work packages, or a shift in billing scale requiring updated contractual justification.

\paragraph{Supplier C: Benign Trimodal Concentration}
Supplier C exhibits a trimodal payment profile characterised by three closely balanced regimes: $\mu_1 = £911$ ($\pi_1 = 0.31$), $\mu_2 = £1,966$ ($\pi_2 = 0.35$) and $\mu_3 = £426$ ($\pi_3 = 0.34$). While the presence of three modes significantly increases the Modality contribution (57.1\%), the overall risk remains categorised as ``Low.'' This is because Structural Dispersion ($3.7\%$), reflects the close proximity of the regimes and their negligible separation. This in turn suggests a single procurement category with tiered payment levels, rather than distinct operational regimes or extreme outliers.

\begin{figure}[htbp]
    \centering
    \includegraphics[width=\textwidth,height=0.4\textheight,keepaspectratio]{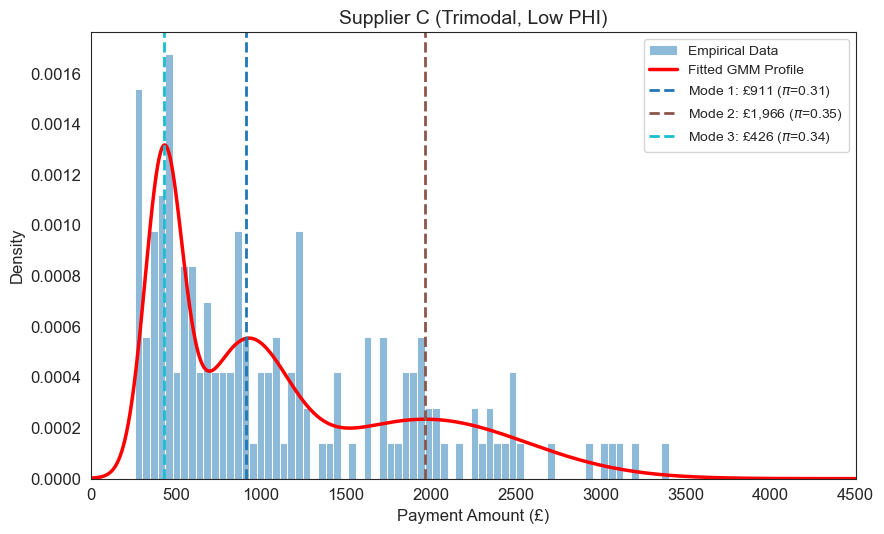}
    \caption{GMM profile for Supplier C. Three closely spaced payment regimes are observed within a narrow financial range, indicating structured but low-separation variation consistent with a low prioritisation profile. Note: The x-axis scale differs across panels to preserve visibility of supplier-specific distributional structure.}
    \label{fig:supplierC}
\end{figure}
\FloatBarrier

An inspection of the categorical detail confirms that, aside from two small-value exceptions, all payments belong to a single \emph{Subjective Detail} category, External Temporary Staff, with the three modes likely corresponding to fixed labour-grade tiers. Because these modes are concentrated within a narrow financial band (approximately between £255 and £3,146) and lack extreme outliers, this pattern is assigned a low prioritisation level. For an auditor, this represents a low-priority signal from a structural complexity perspective: it reflects a stable, structured labour cost model that warrants routine monitoring rather than immediate, detailed investigation.

\paragraph{Supplier D: Quadrimodal Concentration}
Supplier D exhibits four distinct and statistically significant modes ($\pi_i \geq 0.23$), centred at approximately £2,187, £3,105, £3,208, and £3,720. While this high modality contributes 53.3\%, the overall PHI score remains in the ``Moderate'' tier (72.3rd percentile). This is due to the moderating effect of Structural Dispersion ($0.2\%$), which is associated with a relatively narrow spacing between adjacent modes. Indeed, all payments correspond to a single \emph{Subjective Detail} category, Residential Care, indicating that the four modes reflect closely related payment levels within the same service line.

This profile provides an important contrast to Supplier B. While Supplier D is more ``fragmented'' than Supplier B, it is substantially less ``structurally dispersed.'' From an oversight perspective, the PHI avoids over-prioritising this pattern by recognising the multiple modes while correctly accounting for the limited but non-negligible, separation between them. This distinction allows oversight teams to differentiate between granular payment tiers (as seen here) and fundamentally distinct payment scales for the same procurement category identified in Supplier B. Nevertheless, the moderate Tail Behaviour (28.1\%) indicates that the right-most tail near $£4{,}000$ warrants periodic monitoring for emerging outliers.

\begin{figure}[!t]
    \centering
    \includegraphics[width=\textwidth,height=0.4\textheight,keepaspectratio]{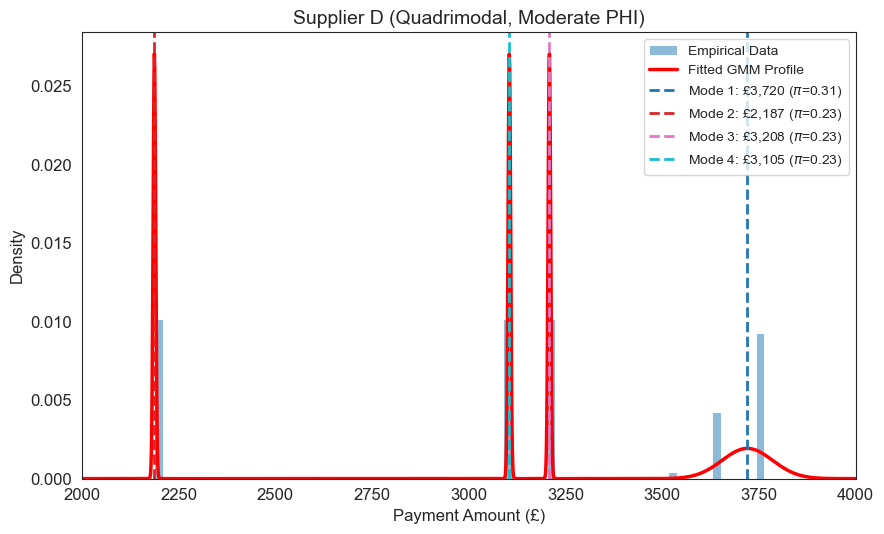}
    \caption{GMM profile for Supplier D. Four distinct but closely aligned payment regimes are identified within a single service category, with minimal separation between modes, resulting in a moderate but not elevated prioritisation level. Note: Due to the very low variance of individual regimes, the fitted density appears sharply peaked, reflecting tightly clustered payment levels rather than model instability. Also, the x-axis scale differs across panels to preserve visibility of supplier-specific distributional structure.}
    \label{fig:supplierD}
\end{figure}
\FloatBarrier

\paragraph{Maximum PHI: Highest-Priority Expert-Review Case}
The supplier with the maximum PHI score represents the highest-priority case for expert review within the analysed cohort. The profile is characterised by an unusually large Structural Dispersion contribution (58.3\%; $D = 27.85$), with the majority of payments clustered at $\mu_1 = £3,088$ ($\pi = 0.81$) and a secondary high-value regime at $\mu_2 = £154,625$ ($\pi = 0.19$). The magnitude of separation between these regimes, combined with elevated Tail Behaviour, indicates a payment structure that is fundamentally atypical relative to the broader supplier population.

\begin{figure}[!t]
    \centering
    \includegraphics[width=\textwidth,height=0.4\textheight,keepaspectratio]{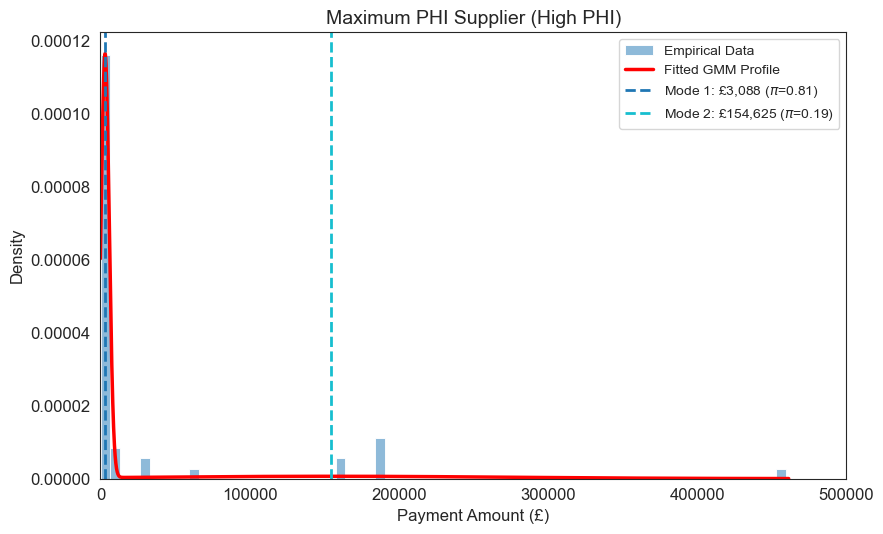}
    \caption{GMM profile for the maximum PHI supplier. A dominant operational regime coexists with a highly separated secondary high-value regime, producing extreme separation and elevated tail behaviour that drive the highest prioritisation score in the sample. Note: The x-axis scale differs across panels to preserve visibility of supplier-specific distributional structure.}
    \label{fig:supplierWorst}
\end{figure}

This structural heterogeneity is reflected in the presence of multiple \emph{Subjective Detail} categories spanning distinct operational functions: Transportation-related expenses, including Bus Services, Subsidised Transport and Bus Passes, coexist alongside administrative and overhead costs such as General Office Expenses, Professional Fees, and Advertising and Publicity. The boxplot visualisation reveals that categories like Other Agencies, Professional Fees, and Service Specific Fees, exhibit pronounced extreme outliers with payment amounts far exceeding typical ranges within those categories, reinforcing the presence of a secondary high-value regime.

From an oversight perspective, this case exemplifies the intended behaviour of the PHI: the joint effect of extreme separation and tail behaviour signals structurally atypical payment dynamics that would reasonably be prioritised for targeted audit review, given this dataset. Such patterns may arise from large one-off contracts, variable pricing schemes, or bundled invoicing practices, which contribute to the multimodal payment distribution and elevated PHI components. Detailed, category-specific investigation, supported by cross-referencing payment data with operational metrics like contract terms and service delivery volumes, is essential to differentiate legitimate operational complexity from irregularities that could include human error or potential wrongdoing. Such a targeted approach enhances the efficiency of resource allocation while facilitating systemic process improvements within high-volume procurement environments.

Taken together, these cases illustrate that the PHI provides a structured and interpretable framework capable of prioritising payment patterns that depart materially from routine operational variation. While the final distinction between legitimate complexity and anomalous behaviour remains a human interpretive task, the index enables audit teams to allocate resources proportionally by focusing investigation where structural signals are strongest. PHI-based prioritisation is relative to the analysed cohort and should be interpreted within the broader institutional and operational context.

\subsection{Visual and Statistical Assessment of Threshold Anchoring}

The preceding supplier-level case studies illustrate PHI's prioritisation behaviour in individual accounts. To further evaluate whether these structural signatures coincide with wider, system-level habits rather than idiosyncratic behaviour, we investigate a phenomenon we term \emph{threshold anchoring}: the propensity for payment regimes to cluster around specific financial benchmarks. Unlike conventional analyses that focus narrowly on avoidance patterns immediately below known limits, our approach captures clustering on either side of an inferred anchor point. 

When administrative thresholds exist, PHI may be aligned directly with thresholds of interest. Because relevant operational thresholds are not always available or recoverable from administrative records, we demonstrate an endogenous peak-extraction procedure as a proof-of-concept for threshold anchoring. The recurring payment levels identified here are not interpreted as exogenous legal or contractual thresholds. Rather, they are empirical anchors visible in the cleaned payment data and used to examine whether high-PHI suppliers interact more strongly with recurring institutional payment levels than lower-PHI suppliers. We emphasise that this characterisation is intentionally descriptive rather than causal; it reflects the practical investigative reality where auditors surface recurring payment levels to identify institutional patterns. These results are presented as descriptive hypotheses for this specific authority, which warrant further validation using independent datasets or contract-level metadata. A fully independent test of threshold anchoring would require either exogenous threshold registers (e.g.\ contract-value schedules or approval-limit tables); this extension is deferred to future work. 

From a management perspective, identifying such clusters is significant because it shifts the focus from individual suppliers to systemic procurement payment habits that may indicate where internal controls and processes may need refinement. By isolating these cohort-wide patterns, decision-makers can better distinguish potential structural risks that may warrant policy-level intervention.

Reference payment peaks were first identified from the empirical distribution of modal centre positions within the high-PHI tier and then examined across the wider cleaned dataset. Specifically, a histogram of high-PHI GMM centre estimates was constructed using £100 bins over the range £0--£30,000, with counts regularised using a Gaussian kernel smoother ($\sigma = 4$ bins). Peaks were extracted using a relative prominence criterion, retaining only those with a prominence exceeding 4\% of the maximum smoothed density and a position above £300 to suppress near-zero boundary artefacts. This procedure identified seven primary payment peaks at £1,150, £7,450, £11,050, £14,350, £16,550, £20,750, and £23,450. Because these peaks are derived endogenously from the high-PHI cohort, the analysis is not presented as an independent confirmatory test of externally known thresholds. Instead, it asks whether these high-PHI-derived anchors also appear in the wider cleaned data, and whether high-PHI suppliers remain more persistently proximate to them, particularly at higher payment values.

Importantly, the existence of recurring payment anchors is not itself treated as anomalous. In public procurement, such anchors may reflect standardised contract values, service tariffs, routine billing conventions, approval practices, or other institutional features. The relevant question is therefore comparative: whether high-PHI suppliers exhibit stronger, more persistent, or higher-value interaction with these shared anchors than lower-PHI suppliers. In this context, the PHI signal does not identify the existence of thresholds per se, but differentiates suppliers by the extent and structure of their interaction with recurring financial reference points.

\begin{figure}[htbp]
    \centering
    \includegraphics[width=\linewidth]{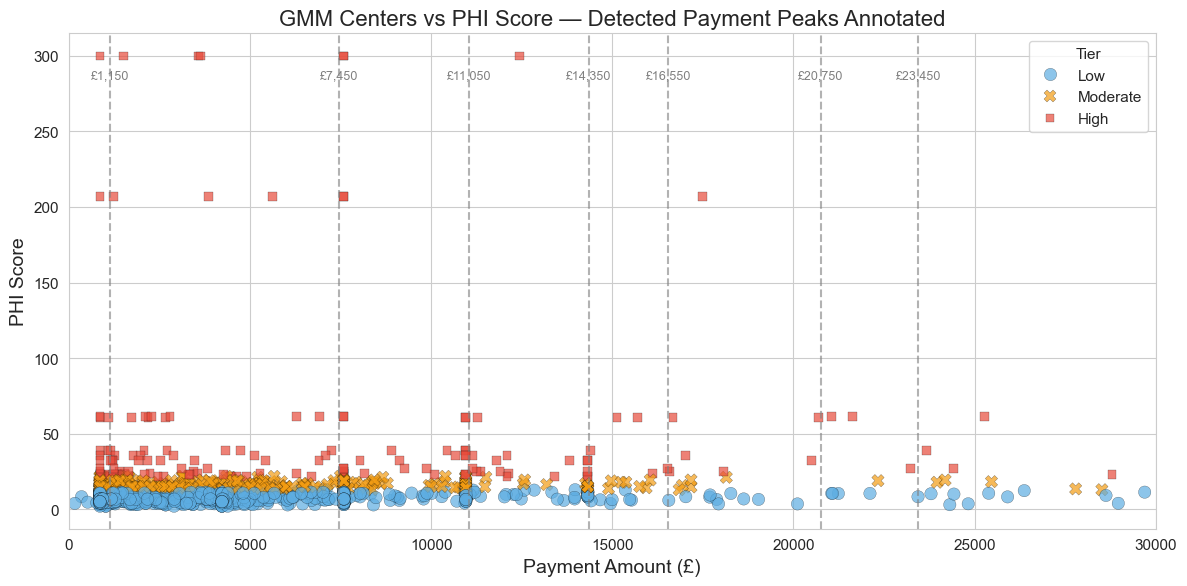}
    \caption{Scatter plot of GMM centre payment amounts versus supplier PHI scores, coloured by PHI tier (Low: blue circles; Moderate: orange crosses; High: red squares). Vertical dashed lines indicate the seven identified empirical reference points at £1,150, £7,450, £11,050, £14,350, £16,550, £20,750, and £23,450. These anchors are visible as recurring payment levels in the cleaned data, but high-PHI suppliers show more persistent vertical alignment across mid- and higher-value anchors, forming a ladder-like pattern of payment regimes. Low-PHI suppliers exhibit clustering primarily at lower-value anchors, consistent with baseline systemic payment regularities.}
    \label{fig:scatter_peaks}
\end{figure}

The relationship between GMM centre positions and PHI scores provides visual insight into threshold-proximate behaviour (Figure~\ref{fig:scatter_peaks}). Because multiple GMM centres may be associated with a single supplier, the analysis is conducted at the level of modal components, with each distinct GMM centre treated as a single observation representing a recurring payment regime within the cohort. Low-PHI suppliers (blue markers) exhibit payment modes primarily concentrated in the lower PHI range, with visible clustering around the lowest-value peaks such as £1,150. In contrast, high-PHI suppliers (red markers) display more persistent vertical alignment across the full range of detected peaks, forming a ladder-like pattern of centres that reflects repeated convergence at shared financial levels across a broad range of PHI scores.

\FloatBarrier

\begin{figure}[htbp]
    \centering
    \includegraphics[width=0.9\linewidth]{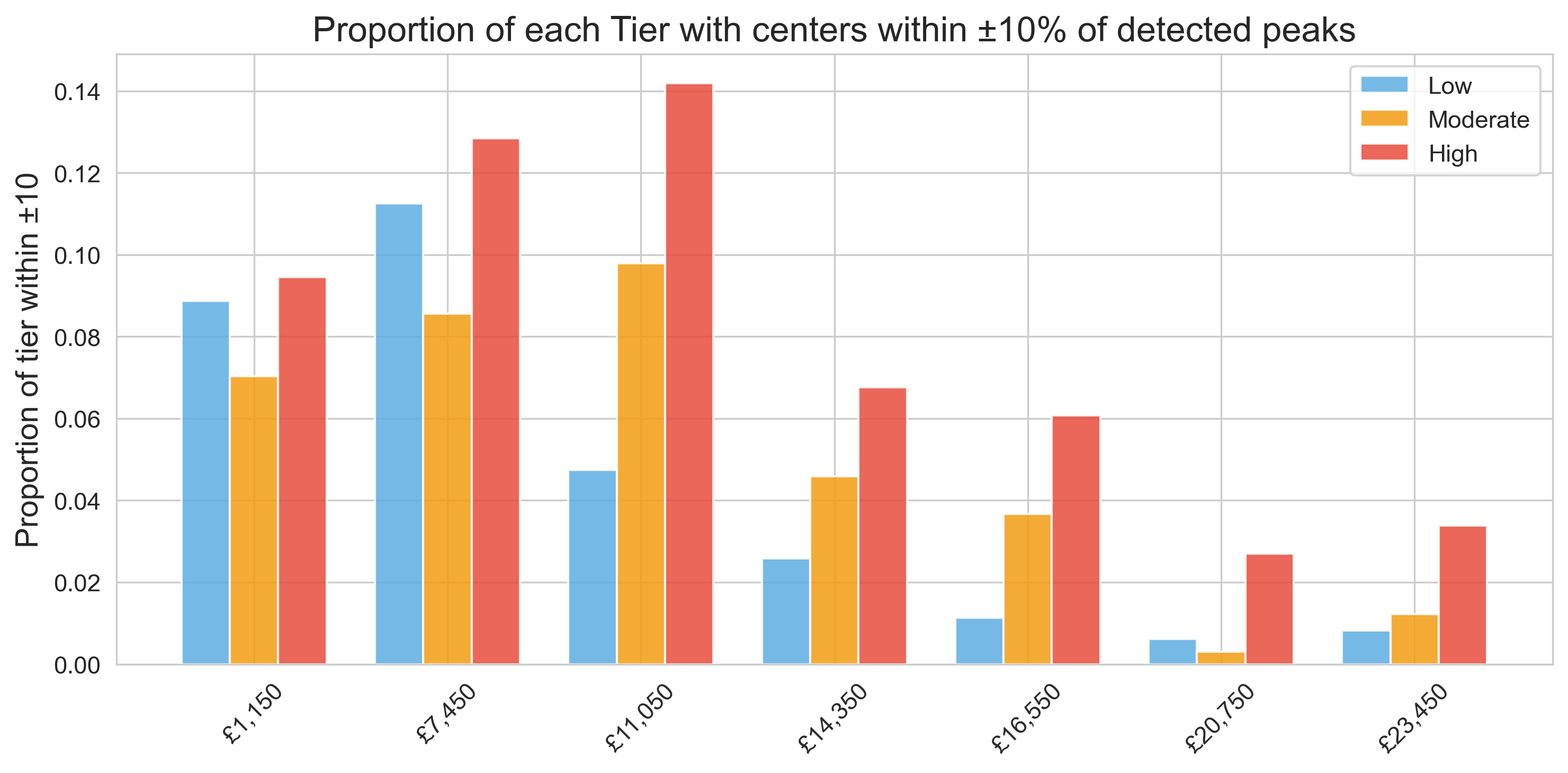}
    \caption{Proportion of each risk tier with GMM centres within ±10\% of detected payment peaks. While all tiers show some proximity at lower-value thresholds, the high-PHI cohort tends to record higher proportions at several mid-to-higher payment levels. Low-PHI proportions are generally lower at these thresholds, but not absent. This pattern is consistent with comparatively stronger threshold anchoring among high-PHI suppliers.}
    \label{fig:peak_props}
\end{figure}

The aforementioned pattern becomes increasingly apparent as payment values rise. To account for differing sample sizes across risk groups, we analysed the proportion of GMM centres within each tier that fall within $\pm 10\%$ of the endogenously detected peaks (Figure~\ref{fig:peak_props}). While all tiers exhibit a degree of proximity at lower-value thresholds, the high-PHI cohort is distinguished by its comparatively higher concentration at mid-to-higher payment levels. Specifically, as payment amounts rise through the £11,050 to £23,450 range, the relative alignment among low-PHI suppliers tends to substantially subside, whereas high-PHI suppliers retain a noticeably greater proportional presence near these thresholds. This suggests that while these benchmarks act as shared institutional reference points across the dataset, the differentiating feature of the high-PHI tier is not merely proximity to anchors in general, but the persistence of proximity at mid-to-higher value anchors.

\FloatBarrier

To assess whether the observed divergence in threshold anchoring between tiers reflects more than a visual pattern, we test whether the prevalence of anchoring in the high-PHI cohort is greater than would be expected under random tier assignment. To this end, we conduct permutation-based tests. For each payment observation, we assign a unique nearest peak by identifying the minimum absolute distance to the set of detected peaks. We then apply a proximity criterion based on a chosen percentage tolerance window (e.g., $\pm 10\%$ of the peak value). Observations falling within this window are classified as `threshold-proximate'. The test statistic is the observed count of high-PHI observations that satisfy this criterion. To generate a null distribution, we hold the payment values and their proximity classifications fixed, while randomly permuting the tier labels ($N=5{,}000$ iterations). For each iteration, we recalculate the count of proximate observations for the newly assigned high tier. The permutation $p$-value is then defined as the proportion of iterations where the permuted count equals or exceeds the observed count.

\begin{figure}[htbp]
    \centering
    \includegraphics[width=0.9\linewidth]{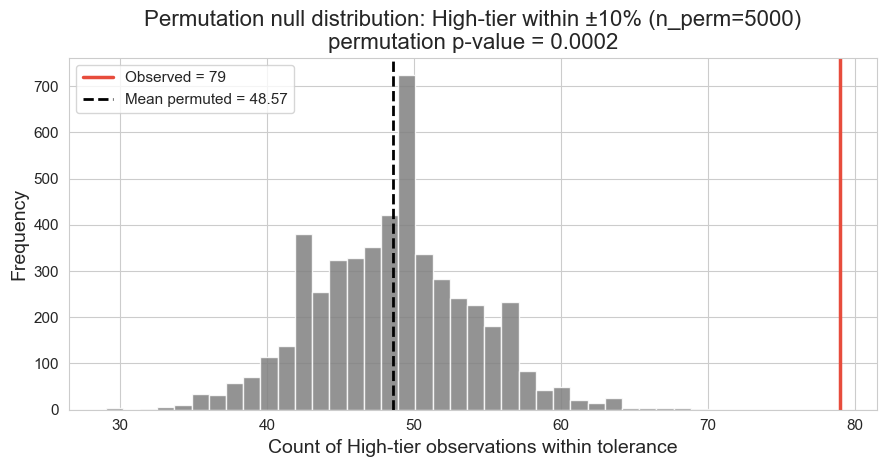}
    \caption{Permutation null distribution of the count of high-tier observations falling within $\pm 10\%$ of detected peaks, based on $N=5{,}000$ random permutations of tier labels. The dashed vertical line indicates the permuted mean (48.6); the red vertical line marks the observed count (79). The one-sided permutation $p$-value is 0.0002. The placement of the observed count in the extreme right tail supports a prioritisation signal for high-PHI suppliers.}
    \label{fig:perm10}
\end{figure}

\begin{figure}[htbp]
    \centering
    \includegraphics[width=0.9\linewidth]{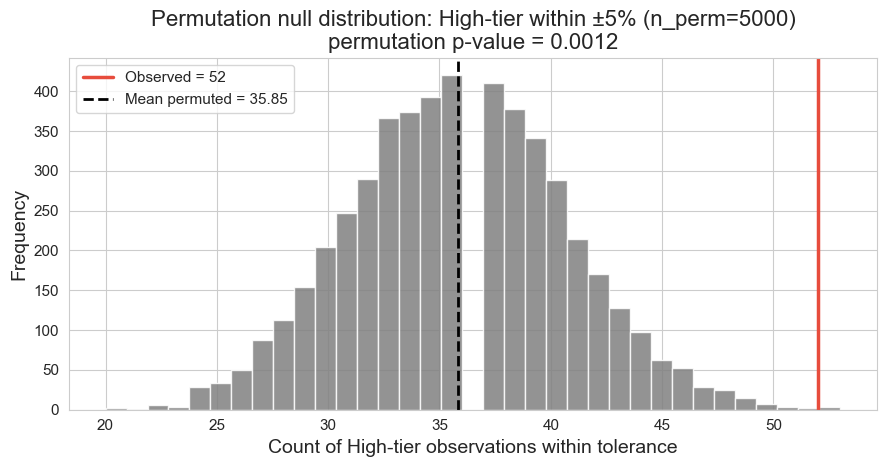}
    \caption{Permutation null distribution of the count of high-tier observations falling within $\pm 5\%$ of detected peaks, based on $N=5{,}000$ random permutations of tier labels. The dashed vertical line indicates the permuted mean (35.9); the red vertical line marks the observed count (52). The one-sided permutation $p$-value is 0.0012. The results remain robust under the stricter tolerance window.}
    \label{fig:perm5}
\end{figure}

High-PHI suppliers demonstrate a 53.4\% prevalence of threshold-proximate behaviour (within $\pm 10\%$ of detected peaks), compared to 29.3\% in the low-PHI cohort. This high-PHI anchoring prevalence (79 observations) significantly exceeds the null expectation of 32.8\% (permuted mean = 48.6; $p = 0.0002$; see Figure~\ref{fig:perm10}). This relationship remains robust under a more stringent $\pm 5\%$ criterion, where the high-tier prevalence of 35.1\% (52 observations) continues to significantly deviate from the null mean of 24.2\% ($p = 0.0012$; see Figure~\ref{fig:perm5}). These results are consistent with high‑PHI suppliers showing a greater propensity to cluster near the inferred payment peaks, particularly the higher‑value peaks, than suppliers in the broader sample. The permutation results therefore support the descriptive claim that PHI is associated with stronger interaction with recurring payment anchors, while leaving causal interpretation and institutional explanation to subsequent audit review.

\FloatBarrier

\FloatBarrier

\begin{figure}[htbp]
    \centering
    \includegraphics[width=0.9\linewidth]{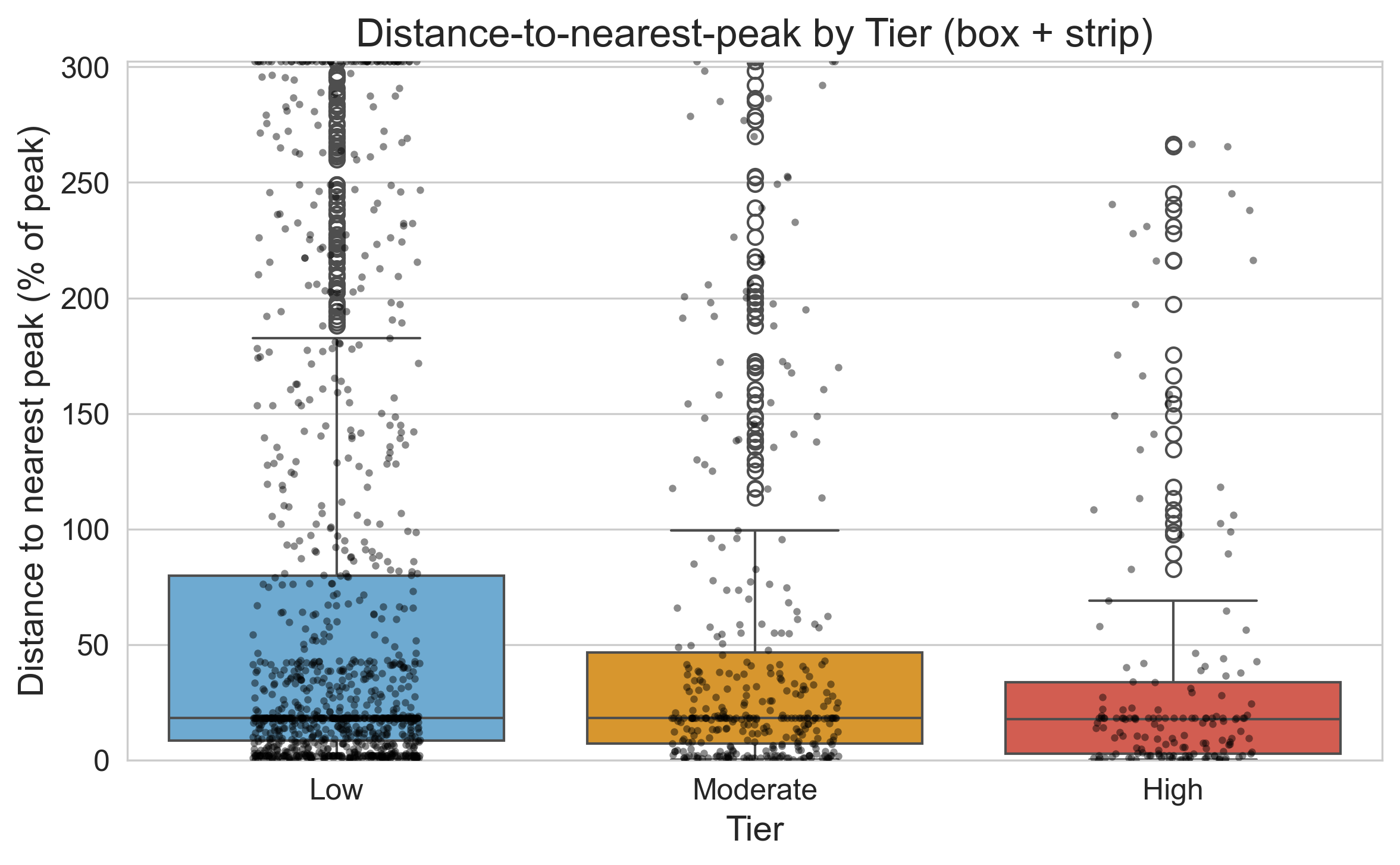}
    \caption{Box‑and‑whisker plot with overlaid strip plot showing the distribution 
    of distance‑to‑nearest‑peak (percentage of the nearest detected peak) by risk tier (unit = GMM centres). Although median distances are similar across tiers, the high‑PHI group exhibits a denser concentration of centres at low distance‑to‑peak values. A two‑sample Kolmogorov--Smirnov test confirms the distributions differ ($K_{\mathrm{S}} = 0.241$, $p = 4.985\times10^{-7}$); see the ECDF comparison in Figure~\ref{fig:cdf_distance}.}
    \label{fig:box_distance}
\end{figure}

\begin{figure}[htbp]
    \centering
    \includegraphics[width=\linewidth]{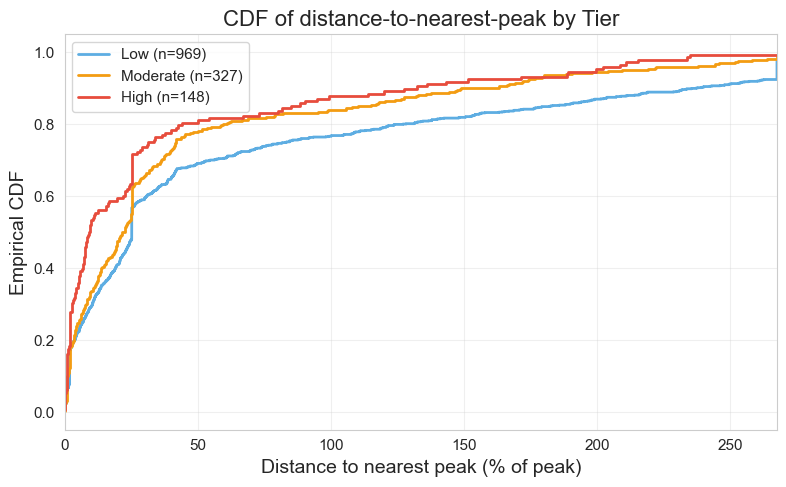}
    \caption{Empirical cumulative distribution functions (ECDFs) of distance‑to‑nearest‑peak (percentage of peak) by risk tier (unit = GMM centres). The high‑PHI curve (red; $n=148$) lies above the low‑PHI curve (blue; $n=969$), indicating that high‑PHI centres accumulate probability mass more rapidly at lower distances (i.e. are more frequently threshold‑proximate). Approximately 41\% of high‑PHI centres fall within $\pm 10\%$ of an inferred peak, compared to 27\% of low‑PHI centres; these proportions mirror the distributional shift confirmed by the K--S test and the excess counts observed in the permutation analysis (Figures~\ref{fig:perm10}--\ref{fig:perm5}).}
    \label{fig:cdf_distance}
\end{figure}

The material distinction between the high‑ and low‑PHI samples is further supported by Figure~\ref{fig:box_distance} and a two‑sample Kolmogorov--Smirnov (K--S) test. While the difference in central tendency between tiers is modest, the empirical cumulative distribution functions (ECDFs; Figure~\ref{fig:cdf_distance}) show that high‑PHI GMM centres accumulate probability mass more rapidly at smaller distances, consistent with a greater concentration of high‑PHI GMM centres in close proximity to inferred payment peaks. The K--S test provides distribution-level evidence that distance-to-nearest-peak (expressed as a percentage of the peak value) differs between the high- and low-PHI groups: the maximum vertical separation between the ECDFs is $K_{\mathrm{S}} = 0.241$ and the associated $p$-value is $p = 4.985\times10^{-7}$.

\FloatBarrier

\FloatBarrier

Combining the visual and statistical evidence suggests that high-PHI suppliers exhibit a distinct modal payment architecture. The permutation tests quantify an excess of high-PHI observations within predefined proximity windows, while the K--S result indicates a broader distributional shift toward inferred payment peaks relative to low-PHI suppliers. Importantly, the visual evidence suggests that this distinction is most pronounced at mid-to-higher value anchors, where low-PHI proximity tends to decline while high-PHI proximity persists. Rather than attributing intent, these findings reinforce positioning PHI as a prioritisation signal: high-PHI payment signatures are more frequently and more materially aligned with recurring payment anchors than those of other tiers. The concordance of a distribution-level shift and excess counts within tolerance windows provides a reasonable basis for targeted review, while remaining exploratory until replicated using independent datasets or contract-level metadata.

As an initial empirical result, this pattern should be examined across additional datasets and institutional settings to establish its potential generality. Additionally, while PHI provides an interpretable prioritisation signal, its wider application for decision support should be complemented by contextual diagnostics such as the threshold-proximity analysis presented here. This multi-layered approach suggests that high-PHI signatures correspond to material structural patterns within a given institutional context, while accounting for variation in unobserved administrative thresholds.

\subsection{Comparison with a Variability Baseline}

A natural question is whether a simpler measure of payment variability could replicate PHI's prioritisation. The coefficient of variation
\[
\mathrm{CV} \;=\; \frac{\sigma}{\mu}
\]
is a straightforward candidate: it is a relative measure, useful for comparing different  datasets in terms of stability, homogeneity or consistency and for analysing variability. To evaluate the relation between CV and PHI, Table~\ref{tab:cv_comparison_top12} presents the Top 12 suppliers by PHI (labels anonymised as S‑01\dots S‑12). For each supplier the table shows the factor values $M$, $A$, $T$, and $D$ together with each factor's contribution  to PHI (according to equation~\eqref{eq:contribution}), and the CV and PHI ranks computed across the analytic cohort.

\begin{table}[htbp]
  \centering
  \caption{Top 12 PHI suppliers, PHI values and contribution, with CV and PHI ranks.}
  \label{tab:cv_comparison_top12}
\resizebox{\textwidth}{!}{%
  \begin{tabular}{@{}c c c c c c c@{}}
    \toprule
    Supplier ID & M & A  & T & D & CV Rank & PHI Rank \\
    \midrule
    S-01 & 2 (12.2\%) & 1.215 (3.4\%) & 4.430 (26.1\%) & 27.853 (58.3\%) & 7 & 1 \\
    S-02 & 2 (13.0\%) & 1.400 (6.3\%) & 2.872 (19.8\%) & 25.730 (60.9\%) & 3 & 2 \\
    S-03 & 2 (16.8\%) & 1.990 (16.7\%) & 2.132 (18.4\%) & 7.250 (48.1\%) & 34 & 3 \\
    S-04 & 3 (26.7\%) & 1.822 (14.6\%) & 2.524 (22.5\%) & 4.408 (36.1\%) & 4 & 4 \\
    S-05 & 3 (30.0\%) & 1.810 (16.2\%) & 3.793 (36.4\%) & 1.897 (17.5\%) & 10 & 5 \\
    S-06 & 3 (30.7\%) & 1.512 (11.5\%) & 2.709 (27.8\%) & 2.917 (29.9\%) & 30 & 6 \\
    S-07 & 4 (39.9\%) & 1.760 (16.3\%) & 2.546 (26.9\%) & 1.796 (16.9\%) & 25 & 7 \\
    S-08 & 2 (21.0\%) & 1.253 (6.8\%) & 4.057 (42.4\%) & 2.675 (29.8\%) & 16 & 8 \\
    S-09 & 2 (21.5\%) & 1.092 (2.7\%) & 3.710 (40.7\%) & 3.087 (35.0\%) & 13 & 9 \\
    S-10 & 2 (21.9\%) & 1.525 (13.3\%) & 2.334 (26.7\%) & 3.348 (38.1\%) & 44 & 10 \\
    S-11 & 3 (34.9\%) & 2.000 (22.0\%) & 3.833 (42.7\%) & 1.011 (0.3\%) & 18 & 11 \\
    S-12 & 4 (45.0\%) & 1.579 (14.8\%) & 2.870 (34.2\%) & 1.205 (6.0\%) & 61 & 12 \\
    \bottomrule
  \end{tabular}
}
  \vspace{2mm}
  \scriptsize\textit{CV Rank and PHI Rank are computed across the analytical sample of  $N=119$ canonical suppliers.}
\end{table}

Across the full cohort of  $N=119$ canonical suppliers, the Spearman rank correlation between CV and PHI is $\rho = 0.310$ ($p = 5.98\times 10^{-4}$), indicating weak monotonic agreement. This suggests partial overlap but not interchangeability: while CV and PHI are correlated to some extent, they rank suppliers differently in ways that matter for prioritisation.

Table~\ref{tab:cv_comparison_top12} clarifies where PHI adds value relative to a CV-only screen. Several suppliers that PHI ranks highly, such as S‑03, S‑06, S‑07 and S‑12, would receive substantially less attention under a CV-only approach and the reasons for these divergences are instructive.

First, CV reports total dispersion but cannot separate between‑regime variance from within‑regime noise. Suppliers with meaningful but not overwhelming regime separation are therefore deprioritised by CV even though PHI elevates them by explicitly quantifying structural heterogeneity. For example, S‑03 (PHI rank 3, CV rank 34) has a significant contribution from structural heterogeneity (D = 48.1\%) that PHI recognises; CV, as a simple moment ratio, treats that signal as ordinary spread and does not reveal its regime‑based origin.

Second, CV conflates modality and tail effects, so it may miss cases where the joint effect of multiple modes and isolated large payments drives investigative interest. S‑12 (PHI rank 12, CV rank 61) illustrates this: the fitted mixture shows three tightly clustered modes plus a relatively isolated and small fourth mode, producing high modal complexity and tail contribution (M = 45.0\%, T = 34.2\%) despite a small between‑cluster variance share (D = 6.0\%). PHI prioritises S‑12 because the combined modality and tail signal indicates compositional complexity and a structural break from the supplier’s typical payment behaviour, that a single $\sigma/\mu$ statistic does not convey.

Third, collapsing variability into one number hides compositional richness and can produce rank inversions: two suppliers with very similar CV values can have markedly different mixes of M, A, T and D. For example, S\nobreakdash-03 and S\nobreakdash-06 have almost identical CVs (1.352 vs 1.377), leading CV to treat them as similarly ranked and not  particularly prominent within the cohort. In contrast, PHI elevates both into the high-priority tier while revealing distinct structural drivers: S\nobreakdash-03 is driven by structural heterogeneity ($D \approx 48\%$), while S\nobreakdash-06 is driven by modal complexity ($M \approx 31\%$). PHI therefore discriminates based on the full multidimensional richness of the data structure, surfacing potential risks that are otherwise collapsed by simpler variability measures.

We therefore offer a conservative interpretation: PHI and CV partly agree, but PHI provides prioritisation leads that CV cannot, particularly when multiple factors such as regime separation or combined modal/tail complexity synergise. CV remains a fast and useful initial screening statistic for broad triage; PHI is most valuable when the analytic objective is to discriminate cases driven by structural regime separation, modal complexity, or atypical tails from those driven by homogeneous spread or isolated outliers. That decomposition yields actionable leads for human-led drill-down and interpretable, testable hypotheses. A more comprehensive evaluation against alternative prioritisation frameworks is an appropriate direction for future work.

\subsection{Sectoral Risk Distribution and Management Prioritisation Triage}

\begin{figure}[!h]
  \centering
  \includegraphics[width=\textwidth]{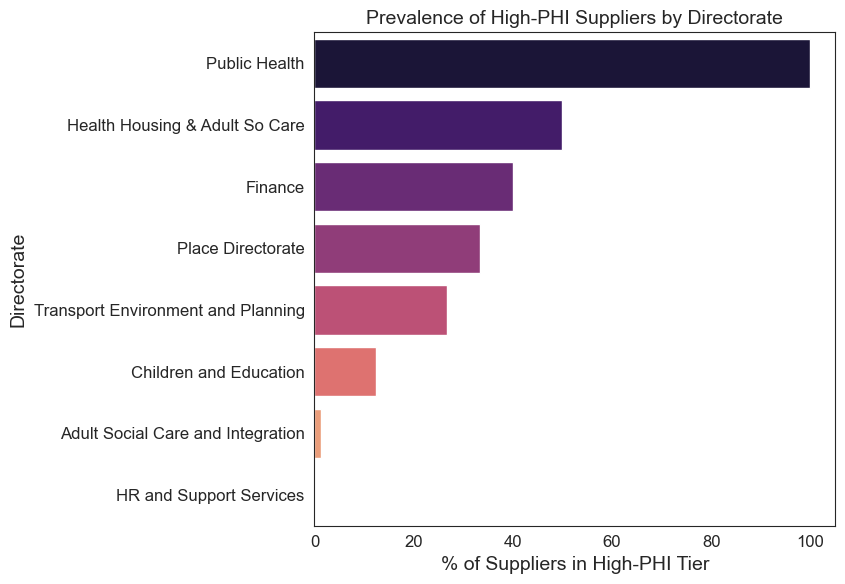}
  \caption{Prevalence (proportion of each directorate) and absolute count of high‑PHI suppliers by directorate (unit = suppliers; filtered sample $N=119$). Small directorates (e.g. Public Health, $n=1$; Finance, $n=5$) should be interpreted cautiously; these bars indicate specific complex accounts rather than systemic sectoral trends. Absolute counts are given in Table~\ref{tab:phi_by_directorate}.}
  \label{fig:high_phi_prevalence}
\end{figure}

Beyond its application in audit triage for individual suppliers, the PHI supports strategic management oversight by identifying where operational complexity is most concentrated.  Applied to the canonical dataset of 1,896 suppliers (filtered to 119 high‑volume vendors with $n\ge 50$), the index isolates 12 high-PHI suppliers, which is roughly 0.6\% of all suppliers and 10.1\% of the high‑volume cohort, providing a focused shortlist for managerial attention.  Mapping these high‑PHI cases to directorates highlights procurement environments that concentrate payment complexity and where targeted human review is likely to be most effective (Figure~\ref{fig:high_phi_prevalence}).

Key observations from this sectoral aggregation include:

\begin{itemize}
    \item \textbf{Sectoral Concentration:} High-PHI signatures are unevenly distributed across the organisation. The \textit{Adult Social Care and Integration} directorate, despite having the largest vendor base ($N=70$), contains only one High-PHI supplier (1.4\%), indicating largely standardised payment patterns. In contrast, smaller and more specialised areas contain a disproportionate share of complex cases. For example, \textit{Finance} has 2 High-PHI suppliers out of 5 vendors, and the \textit{Place Directorate} has 2 out of 6. Additionally, individual High-PHI cases appear in very small directorates such as \textit{Public Health} (1 out of 1) and \textit{Health, Housing \& Adult Social Care} (1 out of 2), though these figures reflect very limited sample sizes. Overall, this pattern suggests that operational complexity is concentrated in a small number of specialised sectors rather than in the largest service areas.
    
    \item \textbf{Resource Concentration:} One-third of High-PHI vendors are within the \textit{Transport, Environment and Planning} directorate ($n=4$), indicating that infrastructure-heavy sectors are more prone to complex payment architectures. This is consistent with the structural profile observed in the maximum PHI case.
    
    \item \textbf{Administrative Triage:} The PHI reveals that 98.6\% of suppliers in \textit{Adult Social Care} follow standardised billing patterns, supporting a lower audit priority for this high-volume, low-complexity sector.
\end{itemize}

\FloatBarrier

\begin{table}[htbp]
  \centering
  \caption{Distribution of PHI tiers by directorate (filtered sample).}
  \label{tab:phi_by_directorate}
  \begin{tabular}{@{}l r r r r@{}}
    \toprule
    Directorate & \multicolumn{1}{c}{High} & \multicolumn{1}{c}{Low} & \multicolumn{1}{c}{Moderate} & \multicolumn{1}{c}{Total} \\
    \midrule
    Adult Social Care and Integration & 1  & 54 & 15 & 70 \\
    Children and Education            & 1  & 6  & 1  & 8  \\
    Finance                           & 2  & 2  & 1  & 5  \\
    HR and Support Services           & 0  & 3  & 1  & 4  \\
    Health Housing \& Adult So Care   & 1  & 1  & 0  & 2  \\
    Housing and Communities           & 0  & 5  & 2  & 7  \\
    Housing and Communities HRA       & 0  & 1  & 0  & 1  \\
    Place Directorate                 & 2  & 3  & 1  & 6  \\
    Public Health                     & 1  & 0  & 0  & 1  \\
    Transport Environment and Planning & 4 & 8  & 3  & 15 \\
    \midrule
    \textbf{Total (filtered sample)}  & \textbf{12} & \textbf{83} & \textbf{24} & \textbf{119} \\
    \bottomrule
  \end{tabular}

  \vspace{2mm}
  \scriptsize\textit{Note: Tiers based on the filtered cohort of high-volume suppliers ($n \ge 50$ payments).}
\end{table}

\FloatBarrier

These sectoral aggregations demonstrate the PHI's precision: rather than implying broad statistical trends, they identify specific complex accounts and furnish management with a fact-based map of payment heterogeneity. This in turn, supports a graduated oversight model, ensuring that interventions, such as focused contract reviews and enhanced controls, are targeted specifically at sectors associated with the highest structural complexity. Given evidence that operational complexity is a key driver of inefficiencies in public procurement \citep{Bandiera2009}, such targeted interventions may yield benefits at multiple levels.

\subsection{Drivers of Structural Heterogeneity: Implications for Decision Support}

To connect underlying patterns with operational supplier behaviour, we examine how the PHI components: modality ($M$), asymmetry ($A$), tail behaviour ($T$), and structural dispersion ($D$), contribute to the index. We quantify each component's log-contribution across the cohort to link statistical signal to actionable hypotheses.

\begin{figure}[!ht]
    \centering
    \includegraphics[width=\textwidth]{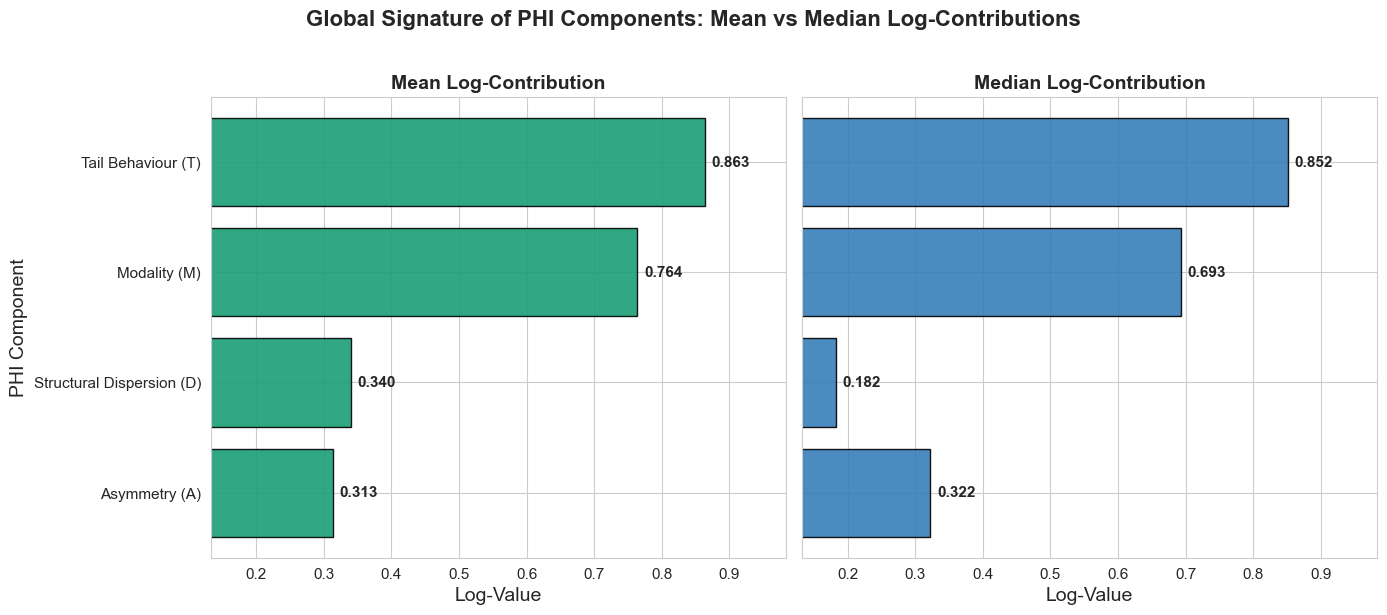}
    \caption{Mean and median log-contributions of PHI components across 
    the full supplier cohort ($N=119$). The consistency between 
    aggregation methods confirms that tail behaviour ($T$) and modality ($M$) 
    are the dominant structural drivers of PHI, with the pattern robust 
    to the influence of extreme observations.}
    \label{fig:global_mean_median}
\end{figure}

\begin{figure}[htbp]
    \centering
    \includegraphics[width=0.66\linewidth]{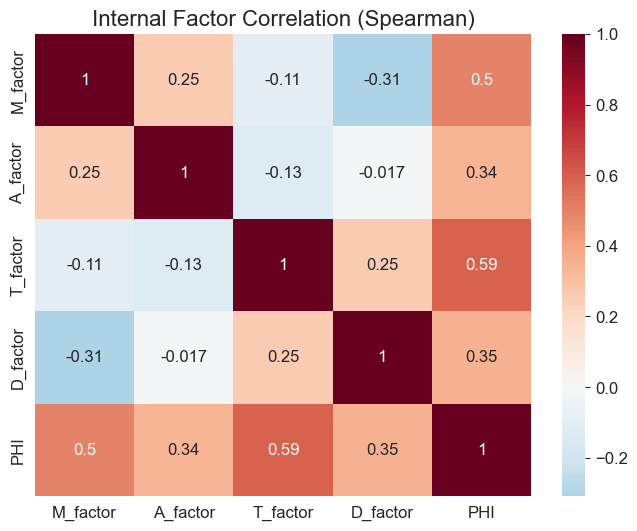}
    \caption{Spearman correlation matrix of PHI components and overall 
    score. Tail Behaviour and modality exhibit the strongest associations 
    with PHI.}
    \label{fig:spearman_heatmap}
\end{figure}

\paragraph{What drives PHI (cohort-level)}
Figure~\ref{fig:global_mean_median} reports mean and median log-contributions across the supplier cohort ($N=119$). The close alignment of mean and median indicates a stable cohort-wide structure rather than dominance by outliers. Tail behaviour ($T$) is the largest contributor (mean = 0.863, median = 0.852), followed by modality ($M$; mean = 0.764, median = 0.693); asymmetry ($A$) and dispersion ($D$) are smaller on average. Structural dispersion shows greater mean–median divergence (mean = 0.340, median = 0.182), signalling a minority of suppliers with substantially larger inter-regime distances. Spearman correlations (Figure~\ref{fig:spearman_heatmap}) reinforce this pattern: PHI correlates moderately with $T$ ($\rho=0.59$) and $M$ ($\rho=0.50$), while associations with $A$ and $D$ are weaker. In sum, payment-amount-concentration and multi-modal structure are primary cohort-level drivers, but no single component fully explains PHI.

\FloatBarrier

\begin{figure}[htbp]
    \centering
    \includegraphics[width=1.1\textwidth]{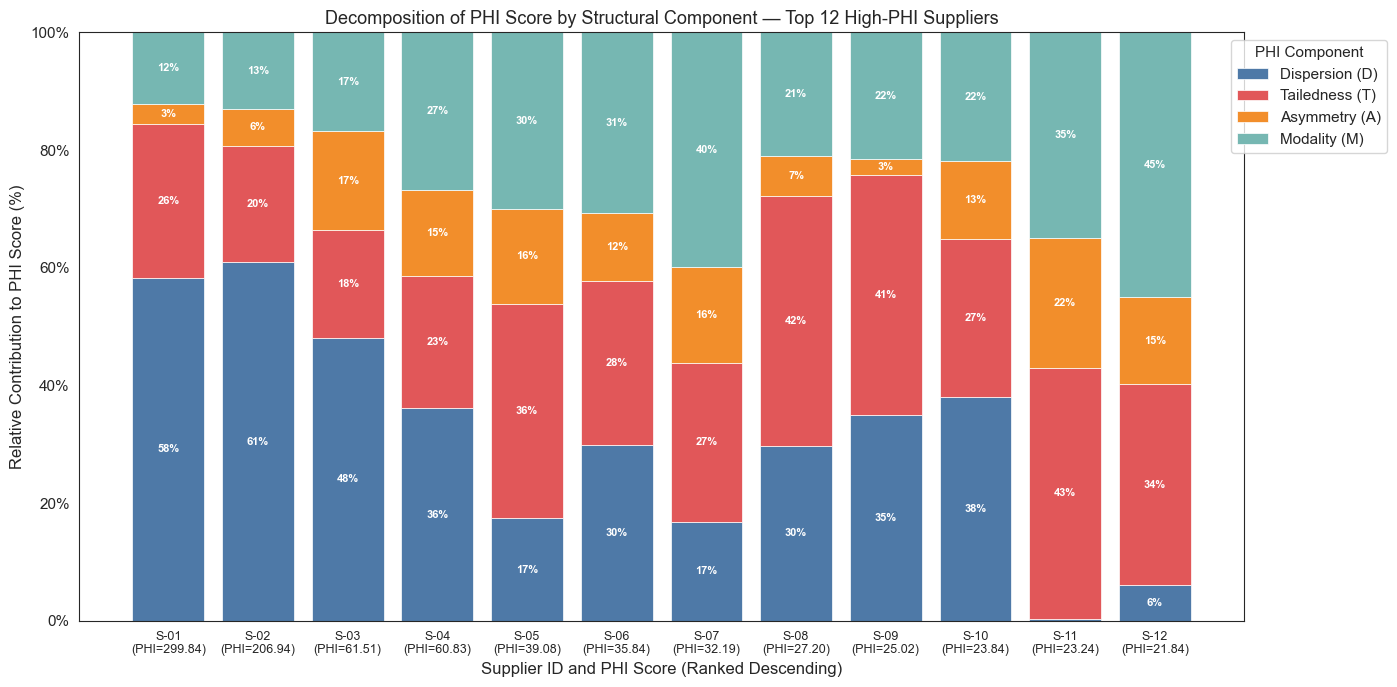}
    \caption{PHI structural driver decomposition for the top 12 suppliers, expressed as percentage contributions of each component. While all entities exhibit high PHI scores, their structural signatures differ substantially. Dispersion ($D$) and tail behaviour ($T$) frequently dominate, but their relative importance varies across suppliers, indicating that similar scores can arise from distinct underlying patterns.}
    \label{fig:top12_decomp}
\end{figure}

\paragraph{Supplier-level decomposition and heterogeneity}
To capture interactions among components we decompose each supplier's PHI into percentage contributions of $M$, $A$, $T$, and $D$. Figure~\ref{fig:top12_decomp} shows that top-ranked suppliers can reach similar PHI magnitudes via different structural signatures: some (e.g., \textbf{S-01}) are dominated by structural dispersion ($D$), others (e.g., \textbf{S-11}) by extreme tail behaviour ($T$), and many present mixed signatures. This heterogeneity accounts for the mean–median divergence: the cohort baseline is stable, while the top tier comprises distinct operational fingerprints rather than scaled versions of a single pattern.

\FloatBarrier

\FloatBarrier

\paragraph{Different Risk Profiles, Not Just Higher Scores}
High PHI scores do not emerge from a uniform structural pattern. As shown in Figure~\ref{fig:top12_decomp}, the highest-ranked suppliers exhibit different structural signatures. This heterogeneity clarifies the mean--median divergence noted in Figure~\ref{fig:global_mean_median}: while the cohort baseline remains stable, the top tier is comprised of unique structural fingerprints rather than amplified versions of the same pattern. 

For example, \textbf{S-01} is defined by widely separated payment regimes (predominantly structural dispersion, $D$), whereas \textbf{S-11} is driven by extreme concentration within regimes (predominantly tail behaviour, $T$). Many other cases reflect complex, multi-factor signatures where no single component dominates. This confirms a core functional property of the PHI: it identifies high-priority entities with similar overall magnitudes but distinct, operationally divergent behaviours. This result is particularly important for evaluating the contribution of $D$. If PHI were driven only by modality or tail behaviour, then high-ranked suppliers would primarily represent either many-mode distributions or extreme-value cases. Instead, the decomposition shows that structural dispersion can dominate the highest-priority cases when payment regimes are materially separated. This supports the intended role of $D$ as a regime-architecture measure that captures a different signal from simple multimodality, spend volume, or tail extremity.

\paragraph{Decision support and interpretable rules}
Because PHI is decomposable, it serves as a diagnostic aid rather than a black box. Dominant components can be mapped to concrete investigative hypotheses (for example, high $T$ suggests clustered pricing or repeated invoicing; high $D$ might suggest split services or mixed price lists), enabling targeted triage. The interpretability of the index is further enhanced by the ability to 'drill down' into specific factors, such as Structural Dispersion, to provide even more precise forensic interpretations and hypotheses. Rather than fixed automated thresholds, analysts can combine component contributions with domain knowledge to construct context-aware, human-in-the-loop rules for prioritisation.

In sum, PHI behaves as intended: elevated scores arise from differing combinations of structural characteristics that can be disentangled via logarithmic decomposition, supporting prioritisation with interpretable rationale. Future work will formally assess the predictive value of these structural drivers and evaluate automated rules for audit prioritisation.

\section{Discussion}
\label{sec:discussion}

The application of the Payment Heterogeneity Index (PHI) to the City of York dataset provides evidence that quantifying structural complexity in payment behaviour is feasible in an interpretable and financially material way within public procurement flows. The top decile of structurally heterogeneous suppliers accounts for nearly a quarter of expenditure in the analytical sample, and PHI isolates roughly 0.6\% of vendors as a high-priority subset for expert review. The index is explicitly designed for high-frequency entities ($n \geq 50$), where multimodal and tail structure can be estimated with reasonable stability. PHI is an unsupervised, decomposable measure that operates effectively with pragmatic default settings; organisation-specific sensitivity analyses and pilot testing would further establish its operational performance, audit-efficiency benefits, and generalisability across institutional settings.

A central property of PHI is its interpretability. By decomposing the overall score into modality ($M$), asymmetry ($A$), tail behaviour ($T$) and structural dispersion ($D$), the framework provides component-level explanations that correspond to specific investigative hypotheses. The evidence suggests that elevated PHI scores are not the result of a single homogeneous pattern; rather, high-priority suppliers exhibit distinct combinations of structural features. This implies that similar scores can emerge from fundamentally different underlying behaviours such as concentrated payment clusters versus widely separated payment regimes. Consequently, the PHI does not merely rank suppliers; it structures the prioritisation rationale, allowing analysts to identify the dominant structural driver before human adjudication commences.

From an operational perspective the interpretability of PHI reduces cognitive load for investigators and accelerates triage. Instead of confronting long lists of opaque alerts, oversight teams receive small sets of high-exposure suppliers together with componentised signatures that suggest concrete lines of enquiry (e.g., repeated invoicing implied by high $T$, or fragmented service lines implied by high $M$ and $D$). This auditable rationale both streamlines manual review and supports more defensible allocation of scarce audit resources. By ensuring algorithmic transparency, PHI addresses the growing need for digital governance, which is a key requirement for decision-support in the public sector \citep{Wachter2017, Janssen2017}. Such interpretable indicators directly support managerial efforts to maintain procurement integrity in high-volume environments \citep{Locatelli2017}, aligning technical innovation with sectoral constraints and governance requirements \citep{Chen2025}.

PHI complements established statistical tools such as Benford-based tests by focusing on realised payment structure rather than digit-level anomalies or tender-stage irregularities. Where Benford-style methods interrogate expected leading-digit frequencies, PHI characterises multimodal regimes and the relative configuration of payment amounts, shifting the analytic lens to post-award expenditure and distributional structure. In this way PHI extends prior procurement risk detection work \citep{Fazekas2016, FazekasKocsis2020} and the broader anomaly-detection literature \citep{Chandola2009} by providing a structural, interpretable signal tailored to transaction-level monitoring.

Because confirmed fraud and error labels are often unavailable in procurement datasets, the evaluation must focus on the utility of the PHI as a decision-support tool rather than as a standard binary classifier. In procurement payment settings, confirmed fraud, error, or misconduct labels are often unavailable, incomplete, or unsuitable as ground truth. PHI is therefore evaluated as an interpretable prioritisation signal: whether it produces financially material, expert-reviewable cases and whether its components behave consistently with the intended triage logic. In this respect, PHI complements rather than replaces digit-based, rule-based, or contract-specific controls.

Beyond procurement, the SHI (of which the PHI is a domain-specific instantiation) offers a broadly applicable, first-principles diagnostic for structural anomalies in empirical data. Empirical measurements across disciplines produce sample distributions, and many oversight problems require distinguishing routine distributional structure from configurations that are unusually multimodal, asymmetric, heavy-tailed, or separated into materially distinct regimes. A composite statistic that quantifies such departures through measures of multimodality, tail behaviour, asymmetry, and structural dispersion provides a principled means to flag processes that are structurally atypical and warrant contextual follow-up. In practice, SHI can be parameterised to serve different operational objectives: PHI, as configured here, targets high-frequency monetary flows using mixture-modelling of payment amounts and a log-decomposition of component contributions. Alternative instantiations may alter component definitions, model-selection criteria, frequency thresholds, or enrichment inputs to prioritise desired signals in domains such as accounts payable analytics, industrial control systems, cybersecurity and more. Framing PHI as a specialised SHI thus both clarifies its theoretical foundation and emphasises its adaptability to situational goals and governance strategies.

For PHI to deliver its full operational value it should be embedded within a broader decision-support architecture. Practical deployment benefits from automated data integration and harmonisation (including supplier-identity resolution and rolling-window profile updates), contextual enrichment that joins PHI outputs to contract metadata and historical audit outcomes, and action-routing workflows that translate component signatures into operational risk tiers while recording investigator feedback. Integrating investigator responses over time will allow PHI to evolve from a purely unsupervised alerting mechanism toward semi-supervised prioritisation that leverages supervisory signals.

Crucially, this integration facilitates a strategic feedback loop between the analytical model and the procurement manager. By systematically recording investigator adjudications on high-PHI alerts, organisations can generate the 'ground truth' labels that are currently missing from administrative datasets. Over time, this allows the framework to evolve from a static discovery tool into a dynamic, learning decision-support system, where the weighting of PHI components can be calibrated against confirmed institutional outcomes. Such a shift from purely unsupervised screening to a semi-supervised workflow ensures that the framework remains resilient to emerging patterns of administrative drift or sophisticated irregularity, effectively personalising the prioritisation logic to the specific risk appetite and operational constraints of the procuring body.

\paragraph{Limitations and Scope} The empirical evaluation is based on a single UK local authority over a six-month window; generalisability to other jurisdictions, authority sizes, or longer fiscal cycles remains to be established. The present study also identifies boundary conditions for PHI's use. PHI's reliability depends on transaction frequency and preprocessing choices: small-sample suppliers ($n<50$) cannot be robustly scored, and mixture-model decisions (e.g., component selection, initialisation) and data-cleaning heuristics influence component estimates. The index is an observational diagnostic, not a causal test: patterns such as repeated alignment at shared payment levels are suggestive but not proof of misconduct. The expert review reported in this paper provides forensic plausibility assessment, not independent adjudication by the procuring authority or a completed audit process. Careful operational governance and human adjudication remain essential to distinguish benign operational complexity from substantive governance risks. In this sense, the effectiveness of PHI as a decision-support tool depends on the capacity of the procuring organisation to interpret and act on its signals, consistent with evidence that bureaucratic competence and procurement governance materially shape procurement outcomes \citep{Decarolis2020}. More broadly, the need to combine data-driven prioritisation with human oversight reflects the well-established trade-off between discretion, efficiency, and abuse in public procurement \citep{Bandiera2021}: while PHI provides the efficiency of targeted screening, human adjudication preserves the necessary administrative discretion to account for context-specific operational complexity.

Future work should pursue concrete validation steps: replication on independent procurement datasets, simulation-based stress tests to probe sensitivity to model choices, and, where institutional records permit, evaluations that link PHI signatures to confirmed audit outcomes using standard metrics such as precision, recall, and area under the precision--recall curve. Because labelled outcomes are often unavailable in this domain, future studies should also consider expert-panel evaluation, blinded case review, and user-centred trials measuring whether decomposed PHI signals reduce investigator workload or improve prioritisation consistency. There is also scope to use large language models to assist data harmonisation and contextual interpretation, further bridging the gap between quantitative risk signals and qualitative administrative oversight..

\section{Conclusions}
\label{sec:conclusion}

This paper introduced the Structural Heterogeneity Index (SHI) and its specialised instantiation, the Payment Heterogeneity Index (PHI), as an interpretable, unsupervised framework for analysing high-frequency post-award procurement payment behaviour. Combining mixture modelling with empirical shape statistics yields a decomposable index that prioritises suppliers and suggests diagnostic hypotheses for expert review. The central technical contribution is the definition of decomposable composite statistics for structural heterogeneity, especially the structural dispersion component, which distinguishes materially separated payment regimes from benign closely spaced multimodal payment tiers. Applied to a municipal dataset, PHI isolates a small set of financially material suppliers using pragmatic default settings and produces interpretable cases consistent with forensic accounting review. The findings provide proof-of-concept for an unsupervised discovery and prioritisation framework i.e they constitute a signal for further investigation rather than primary evidence of misconduct or verified audit findings. While PHI is not a substitute for formal audit, it provides a practical mechanism for converting transaction-level complexity into actionable signals for human-in-the-loop\citep{Mosquera2022, Zheng2017} investigation. Future validation, sensitivity analysis, and integration with contextual data will be critical steps toward embedding PHI within operational decision-support systems for integrity and risk monitoring.

\section*{Declaration of Competing Interests}
Kyriakos Christodoulides is the Managing Founder at Novel Intelligence. The author declares no other competing interests. The author maintained full control over the study design, analysis, and decision to publish, and the research received no specific commercial funding or contract.

\section*{Declaration of Generative AI and AI-assisted Technologies in the Writing Process}
During the preparation of this work, the author used 
\textit{Gemini 3.0 Flash} and \textit{GPT-5.4 mini} (both accessed via the Abacus.ai ChatLLM platform), to improve language and readability. The author reviewed and edited the content as needed and takes full responsibility for the published content.

\section*{Funding}
This research received no specific grant from any funding agency in the public, commercial, or not-for-profit sectors.

\section*{Data Availability}
The data analysed in this study are publicly available from the City of York Council transparency portal (\href{https://data.yorkopendata.org/dataset/all-payments-to-suppliers/resource/87260305-e43a-40a4-866e-ef5e050a3426}{access link}).

\bibliographystyle{elsarticle-harv}
\bibliography{references}

@book{OECD2016,
  author       = {{OECD}},
  year         = {2016},
  title        = {Preventing Corruption in Public Procurement},
  publisher    = {OECD Publishing},
  address      = {Paris},
  doi          = {10.1787/9789264265820-en}
}

@book{OECD2019,
  author       = {{OECD}},
  year         = {2019},
  title        = {Government at a Glance 2019},
  publisher    = {OECD Publishing},
  address      = {Paris},
  doi          = {10.1787/8ccf5c38-en}
}

@article{Fazekas2016,
  author       = {Fazekas, Mihály and Tóth, István János and King, Lawrence P.},
  year         = {2016},
  title        = {Anatomy of Grand Corruption: A Composite Corruption Risk Index Based on Objective Data},
  journal      = {European Journal on Criminal Policy and Research},
  volume       = {22},
  number       = {3},
  pages        = {369--397},
  doi          = {10.1007/s10610-016-9301-z}
}

@article{FazekasKocsis2020,
  author       = {Fazekas, Mihály and Kocsis, Gábor},
  year         = {2020},
  title        = {Uncovering High-Level Corruption: Cross-National Objective Corruption Risk Indicators Using Public Procurement Data},
  journal      = {British Journal of Political Science},
  volume       = {50},
  number       = {1},
  pages        = {155--164},
  doi          = {10.1017/S0007123417000461}
}

@article{Chandola2009,
  author       = {Chandola, Varun and Banerjee, Arindam and Kumar, Vipin},
  year         = {2009},
  title        = {Anomaly Detection: A Survey},
  journal      = {ACM Computing Surveys},
  volume       = {41},
  number       = {3},
  pages        = {15:1--15:58},
  doi          = {10.1145/1541880.1541882}
}

@inproceedings{Carvalho2024,
  author       = {Tyska Carvalho, Jônata and Castro, Márcio and Machado dos Santos, Matheus
                  and Ferrão, Lívia and Schmitz, Fernando Augusto},
  year         = {2024},
  title        = {Detecting Fraud in Public Procurement: A {GMM}-Based Approach to Analyzing Tender Data},
  booktitle    = {Anais do XXXIX Simpósio Brasileiro de Banco de Dados (SBBD 2024)},
  pages        = {207--219},
  publisher    = {Sociedade Brasileira de Computação},
  address      = {Florianópolis, SC, Brazil},
  doi          = {10.5753/sbbd.2024.240649}
}

@inproceedings{Nai2022,
  author       = {Nai, Roberto and Sulis, Emilio and Meo, Rosa},
  year         = {2022},
  title        = {Public Procurement Fraud Detection and Artificial Intelligence Techniques: A Literature Review},
  booktitle    = {Companion Proceedings of the 23rd International Conference on Knowledge Engineering and Knowledge Management (EKAW 2022)},
  series       = {CEUR Workshop Proceedings},
  volume       = {3256},
  publisher    = {CEUR-WS.org},
  url          = {https://ceur-ws.org/Vol-3256/km4law4.pdf}
}

@inproceedings{Rabuzin2019,
  author       = {Rabuzin, Kornelije and Modrušan, Nikola},
  year         = {2019},
  title        = {Prediction of Public Procurement Corruption Indices Using Machine Learning Methods},
  booktitle    = {Proceedings of the 11th International Joint Conference on Knowledge Discovery,
                  Knowledge Engineering and Knowledge Management (IC3K 2019) -- Volume 3: KMIS},
  pages        = {333--340},
  publisher    = {SciTePress},
  doi          = {10.5220/0008353603330340}
}

@misc{CityYork2025,
  author       = {{City of York Council}},
  title        = {Payments over {£}250 Dataset, 2025--2026 Fiscal Year},
  year         = {2025},
  url          = {https://data.yorkopendata.org/dataset/all-payments-to-suppliers/resource/87260305-e43a-40a4-866e-ef5e050a3426},
  note         = {Accessed: 2025-12-26}
}

@book{Nigrini2012,
  author       = {Nigrini, Mark J.},
  title        = {Benford's Law: Applications for Forensic Accounting, Auditing, and Fraud Detection},
  year         = {2012},
  publisher    = {John Wiley \& Sons},
  address      = {Hoboken, New Jersey},
  doi          = {10.1002/9781119203094}
}

@article{Dempster1977,
  author       = {Dempster, A. P. and Laird, N. M. and Rubin, D. B.},
  title        = {Maximum Likelihood from Incomplete Data via the {EM} Algorithm},
  journal      = {Journal of the Royal Statistical Society: Series B (Methodological)},
  volume       = {39},
  number       = {1},
  pages        = {1--22},
  year         = {1977},
  doi          = {10.1111/j.2517-6161.1977.tb01600.x}
}

@book{McLachlan2000,
  author       = {McLachlan, Geoffrey J. and Peel, David},
  title        = {Finite Mixture Models},
  publisher    = {John Wiley \& Sons},
  year         = {2000},
  doi          = {10.1002/0471721182}
}

@article{Schwarz1978,
  author       = {Schwarz, Gideon},
  title        = {Estimating the Dimension of a Model},
  journal      = {The Annals of Statistics},
  volume       = {6},
  number       = {2},
  pages        = {461--464},
  year         = {1978},
  doi          = {10.1214/aos/1176344136}
}

@article{Palguta2017,
  author       = {Palguta, Ján and Pertold, Filip},
  title        = {Manipulation of Procurement Contracts: Evidence from the Introduction of Discretionary Thresholds},
  journal      = {American Economic Journal: Economic Policy},
  volume       = {9},
  number       = {2},
  pages        = {293--315},
  year         = {2017},
  doi          = {10.1257/app.20160289}
}

@article{Bandiera2009,
  author       = {Bandiera, Oriana and Prat, Andrea and Valletti, Tommaso},
  title        = {Active and Passive Waste in Government Spending: Evidence from a Policy Experiment},
  journal      = {American Economic Review},
  volume       = {99},
  number       = {4},
  pages        = {1278--1308},
  year         = {2009},
  doi          = {10.1257/aer.99.4.1278}
}

@article{Janssen2012,
  author       = {Janssen, Marijn and Charalabidis, Yannis and Zuiderwijk, Anneke},
  title        = {Benefits, Adoption Barriers and Myths of Open Data and Open Government},
  journal      = {Information Systems Management},
  volume       = {29},
  number       = {4},
  pages        = {258--268},
  year         = {2012},
  doi          = {10.1080/10580530.2012.716740}
}

@article{West2016,
  author       = {West, Jarrod and Bhattacharya, Maumita},
  title        = {Intelligent Financial Fraud Detection: A Survey},
  journal      = {Computers \& Security},
  volume       = {57},
  pages        = {47--66},
  year         = {2016},
  doi          = {10.1016/j.cose.2015.09.005}
}

@article{Ngai2011,
  author       = {Ngai, E. W. T. and Hu, Yong and Wong, Y. H. and Chen, Yijun and Sun, Xin},
  title        = {The Application of Data Mining Techniques in Financial Fraud Detection:
                  A Classification Framework and Academic Review},
  journal      = {Decision Support Systems},
  volume       = {50},
  number       = {3},
  pages        = {559--569},
  year         = {2011},
  doi          = {10.1016/j.dss.2010.08.006}
}

@article{Bertot2010,
  author       = {Bertot, John C. and Jaeger, Paul T. and Grimes, Justin M.},
  title        = {Using {ICTs} to Create a Culture of Transparency: {E}-Government and Social Media
                  as Openness and Anti-Corruption Tools for Societies},
  journal      = {Government Information Quarterly},
  volume       = {27},
  number       = {3},
  pages        = {264--271},
  year         = {2010},
  doi          = {10.1016/j.giq.2010.03.001}
}

@incollection{Spagnolo2012,
  author       = {Spagnolo, Giancarlo},
  title        = {Reputation, Incentives and Corruption in Public Procurement},
  booktitle    = {The Law and Economics of Public Procurement Reform},
  editor       = {Piga, Gustavo and Treumer, Steen},
  publisher    = {Routledge},
  pages        = {311--325},
  year         = {2012}
}

@article{Hodge2004,
  author       = {Hodge, Victoria J. and Austin, Jim},
  title        = {A Survey of Outlier Detection Methodologies},
  journal      = {Artificial Intelligence Review},
  volume       = {22},
  number       = {2},
  pages        = {85--126},
  year         = {2004},
  doi          = {10.1023/B:AIRE.0000045502.10941.a9}
}

@article{Coviello2018,
  author       = {Coviello, Decio and Guglielmo, Andrea and Spagnolo, Giancarlo},
  title        = {The Effect of Discretion on Procurement Performance},
  journal      = {Management Science},
  volume       = {64},
  number       = {2},
  pages        = {715--738},
  year         = {2018},
  doi          = {10.1287/mnsc.2016.2650}
}

@article{Pedregosa2011,
  author       = {Pedregosa, F. and Varoquaux, G. and Gramfort, A. and Michel, V. and
                  Thirion, B. and Grisel, O. and Blondel, M. and Prettenhofer, P. and
                  Weiss, R. and Dubourg, V. and Vanderplas, J. and Passos, A. and
                  Cournapeau, D. and Brucher, M. and Perrot, M. and Duchesnay, E.},
  title        = {Scikit-learn: Machine Learning in {Python}},
  journal      = {Journal of Machine Learning Research},
  volume       = {12},
  pages        = {2825--2830},
  year         = {2011},
  url          = {https://jmlr.org/papers/v12/pedregosa11a.html}
}

@article{Decarolis2020,
  author       = {Decarolis, Francesco and Giuffrida, Leonardo M. and Iossa, Elisabetta and
                  Mollisi, Vincenzo and Spagnolo, Giancarlo},
  title        = {Bureaucratic Competence and Procurement Outcomes},
  journal      = {The Journal of Law, Economics, and Organization},
  volume       = {36},
  number       = {3},
  pages        = {537--597},
  year         = {2020},
  doi          = {10.1093/jleo/ewaa015}
}

@article{Carril2021,
  author       = {Carril, Rodrigo},
  title        = {Rules versus Discretion in Public Procurement},
  journal      = {Journal of Political Economy},
  volume       = {129},
  number       = {9},
  pages        = {2519--2564},
  year         = {2021},
  doi          = {10.1086/714768}
}

@techreport{Bandiera2021,
  author       = {Bandiera, Oriana and Bosio, Erika and Spagnolo, Giancarlo},
  title        = {Discretion, Efficiency, and Abuse in Public Procurement},
  institution  = {Centre for Economic Policy Research (CEPR)},
  year         = {2021},
  number       = {16743},
  type         = {Discussion Paper},
  url          = {https://cepr.org/publications/DP16743}
}

@article{Chen2025,
  author    = {Chen, Yisong and Zhao, Chuqing and Xu, Yixin and Nie, Chuanhao and Zhang, Yixin},
  title     = {Deep Learning in Financial Fraud Detection: Innovations, Challenges, and Applications},
  journal   = {Data Science and Management},
  year      = {2025},
  doi       = {10.1016/j.dsm.2025.08.002},
  note      = {In press},
  publisher = {KeAi Communications Co. Ltd.}
}

@article{Caglayan2022,
  author  = {Caglayan, Mustafa and Talavera, Oleksandr and Zhang, Wei},
  title   = {Public procurement and the supply of public goods},
  journal = {Journal of Economic Behavior \& Organization},
  year    = {2022},
  volume  = {200},
  pages   = {189--212}
}

@article{Wachter2017,
  title={Counterfactual explanations without opening the black box: Automated decisions and the GDPR},
  author={Wachter, Sandra and Mittelstadt, Brent and Russell, Chris},
  journal={Harvard Journal of Law \& Technology},
  volume={31},
  pages={841},
  year={2017}
}

@article{Janssen2017,
  title={Factors influencing big data decision-making quality},
  author={Janssen, Marijn and van der Voort, Haiko and Wahyudi, Agustinus},
  journal={Journal of Business Research},
  volume={70},
  pages={338--345},
  year={2017},
  publisher={Elsevier}
}

@article{Locatelli2017,
  title={Corruption in public projects and megaprojects: There is an elephant in the room!},
  author={Locatelli, Giorgio and Mariani, Giacomo and Sainati, Tristano and Greco, Marco},
  journal={International Journal of Project Management},
  volume={35},
  number={3},
  pages={252--268},
  year={2017},
  publisher={Elsevier}
}

@inproceedings{Mosquera2022,
  title={Human-in-the-Loop Anomaly Detection and Explanation},
  author={Mosquera, Lucas and Adeline, Benoit and de Rochebouet, Anne-Sophie and Lutz, Michel and Mouret, Alexandre},
  booktitle={International Conference on Information Processing and Management of Uncertainty in Knowledge-Based Systems},
  pages={577--589},
  year={2022},
  publisher={Springer}
}

@inproceedings{Zheng2017,
  title={Human-in-the-loop learning-oriented anomaly detection},
  author={Zheng, Panpan and Kao, Shuo-Han and Chu, Chia-Tung and Chen, Yun-Chun and Wang, Wei-Chung and Chien, I-Hsiang},
  booktitle={2017 IEEE International Conference on Big Data (Big Data)},
  pages={2247--2251},
  year={2017},
  publisher={IEEE}
}

\end{document}